\documentclass[twoside,twocolumn,9pt]{article}
\usepackage{extsizes}
\usepackage[super,sort&compress,comma]{natbib} 
\usepackage[version=3]{mhchem}
\usepackage[left=1.5cm, right=1.5cm, top=1.785cm, bottom=2.0cm]{geometry}
\usepackage{balance}
\usepackage{mathptmx}
\usepackage{sectsty}
\usepackage{graphicx} 
\usepackage{lastpage}
\usepackage[format=plain,justification=justified,singlelinecheck=false,font={stretch=1.125,small,sf},labelfont=bf,labelsep=space]{caption}
\usepackage{float}
\usepackage{fancyhdr}
\usepackage{fnpos}
\usepackage[english]{babel}
\addto{\captionsenglish}{%
  \renewcommand{\refname}{Notes and references}
}
\usepackage{array}
\usepackage{droidsans}
\usepackage{charter}
\usepackage[T1]{fontenc}
\usepackage[usenames,dvipsnames]{xcolor}
\usepackage{setspace}
\usepackage[compact]{titlesec}
\usepackage{hyperref}
\usepackage{epstopdf}%This line makes .eps figures into .pdf - please comment out if not required.

\definecolor{cream}{RGB}{222,217,201}

\begin{document}

\pagestyle{fancy}
\thispagestyle{plain}
\fancypagestyle{plain}{
%%%HEADER%%%
\renewcommand{\headrulewidth}{0pt}
}
%%%END OF HEADER%%%

%%%PAGE SETUP - Please do not change any commands within this section%%%
\makeFNbottom
\makeatletter
\renewcommand\LARGE{\@setfontsize\LARGE{15pt}{17}}
\renewcommand\Large{\@setfontsize\Large{12pt}{14}}
\renewcommand\large{\@setfontsize\large{10pt}{12}}
\renewcommand\footnotesize{\@setfontsize\footnotesize{7pt}{10}}
\makeatother

\renewcommand{\thefootnote}{\fnsymbol{footnote}}
\renewcommand\footnoterule{\vspace*{1pt}% 
\color{cream}\hrule width 3.5in height 0.4pt \color{black}\vspace*{5pt}} 
\setcounter{secnumdepth}{5}

\makeatletter 
\renewcommand\@biblabel[1]{#1}
\renewcommand\@makefntext[1]% 
{\noindent\makebox[0pt][r]{\@thefnmark\,}#1}
\makeatother 
\renewcommand{\figurename}{\small{Fig.}~}
\sectionfont{\sffamily\Large}
\subsectionfont{\normalsize}
\subsubsectionfont{\bf}
\setstretch{1.125} %In particular, please do not alter this line.
\setlength{\skip\footins}{0.8cm}
\setlength{\footnotesep}{0.25cm}
\setlength{\jot}{10pt}
\titlespacing*{\section}{0pt}{4pt}{4pt}
\titlespacing*{\subsection}{0pt}{15pt}{1pt}
%%%END OF PAGE SETUP%%%

%%%FOOTER%%%
\fancyfoot{}
\fancyfoot[LO,RE]{\vspace{-7.1pt}\includegraphics[height=9pt]{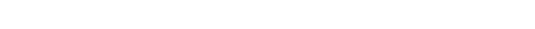}}
\fancyfoot[CO]{\vspace{-7.1pt}\hspace{11.9cm}}%\includegraphics{head_foot/RF}}
\fancyfoot[CE]{\vspace{-7.2pt}\hspace{-13.2cm}}%\includegraphics{head_foot/RF}}
\fancyfoot[RO]{\footnotesize{\sffamily{1--\pageref{LastPage} ~\textbar  \hspace{2pt}\thepage}}}
\fancyfoot[LE]{\footnotesize{\sffamily{\thepage~\textbar\hspace{4.65cm} 1--\pageref{LastPage}}}}
\fancyhead{}
\renewcommand{\headrulewidth}{0pt} 
\renewcommand{\footrulewidth}{0pt}
\setlength{\arrayrulewidth}{1pt}
\setlength{\columnsep}{6.5mm}
\setlength\bibsep{1pt}
%%%END OF FOOTER%%%

%%%FIGURE SETUP - please do not change any commands within this section%%%
\makeatletter 
\newlength{\figrulesep} 
\setlength{\figrulesep}{0.5\textfloatsep} 

\newcommand{\topfigrule}{\vspace*{-1pt}% 
\noindent{\color{cream}\rule[-\figrulesep]{\columnwidth}{1.5pt}} }

\newcommand{\botfigrule}{\vspace*{-2pt}% 
\noindent{\color{cream}\rule[\figrulesep]{\columnwidth}{1.5pt}} }

\newcommand{\dblfigrule}{\vspace*{-1pt}% 
\noindent{\color{cream}\rule[-\figrulesep]{\textwidth}{1.5pt}} }

\makeatother
%%%END OF FIGURE SETUP%%%

%%%TITLE, AUTHORS AND ABSTRACT%%%
\twocolumn[
  \begin{@twocolumnfalse}
%{\includegraphics[height=30pt]{head_foot/PCCP}\hfill\raisebox{0pt}[0pt][0pt]{\includegraphics[height=55pt]{head_foot/RSC_LOGO_CMYK}}\\[1ex]
%\includegraphics[width=18.5cm]{head_foot/header_bar}}\par
%\vspace{1em}
\sffamily
\begin{tabular}{m{4.5cm} p{13.5cm} }

\includegraphics{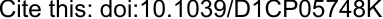} & \noindent\LARGE{\textbf{Photoelectron circular dichroism in angle-resolved photoemission from liquid fenchone$^{\dag}$}} \\%Article title goes here instead of the text "This is the title"
\vspace{0.3cm} & \vspace{0.3cm} \\

 & \noindent\large{Marvin Pohl,\textit{$^{a,b,c,\ddag}$} 
 Sebastian Malerz,\textit{$^{a,\ddag}$}
 Florian Trinter,\textit{$^{a,d}$}
 Chin Lee,\textit{$^{b,c}$}
 Claudia Kolbeck,\textit{$^{a,\S}$}
 Iain Wilkinson,\textit{$^{e}$}
 Stephan Thürmer,\textit{$^{f}$}
 Daniel M.\ Neumark,\textit{$^{b,c}$}
 Laurent Nahon,\textit{$^{g}$}
 Ivan Powis,\textit{$^{h}$}
 Gerard Meijer,\textit{$^{a}$}
 Bernd Winter,\textit{$^{a}$}
 Uwe Hergenhahn$^{\ast}$\textit{$^{a}$}} \\%Author names go here instead of "Full name", etc.

& \noindent\normalsize{We present an experimental X-ray photoelectron circular dichroism (PECD) study of liquid fenchone at the C~1s edge.
A novel setup to enable PECD measurements on a liquid microjet [Malerz \textit{et al., Rev. Sci. Instrum.}, 2022, \textbf{93}, 015101] was used.
For the C~1s line assigned to fenchone's carbonyl carbon, a non-vanishing asymmetry is found in the intensity of photoelectron spectra acquired under a fixed angle in the backward-scattering plane.
This experiment paves the way towards a novel probe of the chirality of organic/biological molecules in aqueous solution.}\\ % Any references in the abstract should be written out in full \textit{e.g.}\ [Surname \textit{et al., Journal Title}, 2000, \textbf{35}, 3523].} \\%The abstrast goes here instead of the text "The abstract should be..."

\end{tabular}

 \end{@twocolumnfalse} \vspace{0.6cm}

  ]
%%%END OF TITLE, AUTHORS AND ABSTRACT%%%

%%%FONT SETUP - please do not change any commands within this section
\renewcommand*\rmdefault{bch}\normalfont\upshape
\rmfamily
\section*{}
\vspace{-1cm}

%%%FOOTNOTES%%%

\footnotetext{\textit{$^{a}$~Molecular Physics, Fritz-Haber-Institut der Max-Planck-Gesellschaft, Faradayweg 4-6, 14195 Berlin, Germany. E-mail: hergenhahn@fhi-berlin.mpg.de}}
\footnotetext{\textit{$^{b}$~Department of Chemistry, University of California, Berkeley, CA 94720, USA}}
\footnotetext{\textit{$^{c}$~Chemical Sciences Division, Lawrence Berkeley National Laboratory, Berkeley, CA 94720, USA}}
\footnotetext{\textit{$^{d}$~Institut für Kernphysik, Goethe-Universität Franfurt am Main, Max-von-Laue-Straße 1, 60438 Frankfurt am Main, Germany}}
\footnotetext{\textit{$^{e}$~Department of Locally-Sensitive \& Time-Resolved Spectroscopy, Helmholtz-Zentrum
Berlin für Materialien und Energie, Hahn-Meitner-Platz 1, 14109 Berlin, Germany}}
\footnotetext{\textit{$^{f}$~Department of Chemistry, Graduate School of Science, Kyoto University,
Kitashirakawa-Oiwakecho, Sakyo-Ku, Kyoto 606-8502, Japan}}
\footnotetext{\textit{$^{g}$~Synchrotron SOLEIL, L’Orme des Merisiers, St. Aubin, BP 48, 91192 Gif sur Yvette, France}}
\footnotetext{\textit{$^{h}$~School of Chemistry, The University of Nottingham, University Park, Nottingham, UK}}

%Please use \dag to cite the ESI in the main text of the article.
%If you article does not have ESI please remove the the \dag symbol from the title and the footnotetext below.
\footnotetext{\dag~Electronic Supplementary Information (ESI) available: Raw data, further details on peak-background separation. See DOI: 10.1039/cXCP00000x/}
%additional addresses can be cited as above using the lower-case letters, c, d, e... If all authors are from the same address, no letter is required

\footnotetext{\ddag~These authors contributed equally to this work.}
\footnotetext{\S~Current address: sonUtec GmbH, Mittlere-Motsch-Straße 26, 96515 Sonneberg, Germany.}

%%%END OF FOOTNOTES%%%

%%%MAIN TEXT%%%%
%\onecolumn   %  UHe - have a more preprint-like style, remove this line before submission
\section{Introduction}
Many of the molecules providing the basis of living matter are chiral, that is they may exist in two different 3-D structural forms, which are mirror images of each other. 
Due to steric effects, these two forms or enantionmers may behave very differently when they interact with other chiral partners: this is chiral recognition, a fundamental metabolic process. Furthermore, the chiral molecular building blocks of terrestrial life, such as amino-acids and nucleic acid sugars, are almost exclusively found as single enantiomers, a fascinating property known as the homochirality of life.\cite{Meierhenrich} 
As a consequence, it is immensely important to have the means to distinguish between enantiomers at the molecular level, despite them possessing largely identical physico-chemical properties such as mass, spectra, and energetics (apart from putative tiny electroweak effects\cite{Letokhov}).
Therefore, chiral discrimination, or recognition, requires interaction with another chiral entity. A common example is the interaction with circularly polarized light (CPL), which gives rise to the well-known circular dichroism (CD) effect in absorption.\cite{Pasteur,CD} 
Relatedly, chiral (spin-polarized) electrons have also been shown to discriminate for the molecular-level handedness of a sample.\cite{Kessler, Zacharias}

It is of great appeal that elements of these two techniques are combined in yet another effect that discriminates between different enantiomers of a species, namely photoelectron circular dichroism (PECD). 
This term designates an asymmetry in the angle-resolved photoelectron (PE) flux after ionization of a sample of chiral molecules with circularly polarized light.
The effect requires a suitable geometry of the experiment, as it vanishes in the plane perpendicular to the photon propagation direction (`dipole plane').
It can be observed as a difference of photoelectron intensity between two measurements, in which either (1) the same sample is probed by left- vs. right-handed circularly polarized light, or (2) the same sample is probed by any helicity of the light, and electrons are collected under two different angles, one in the forward and the other in the backward scattering direction, with the two angles being mirror imaged at the dipole plane, or (3) by probing the two different enantiomers of a substance with either helicity, at an angle outside of the dipole plane.
Historically, the potential existence of PECD was noted in theoretical papers in the 1970s,\cite{Ritchie1,Ritchie2} but only abstract model systems were considered, and these works received only minor attention at that time. 
It was over twenty years later that a dedicated numerical simulation on actual molecules by I.~Powis suggested that this effect could have an observable magnitude.\cite{Powis_2000} In fact, it was simulated to be significantly greater than that of more conventional chiroptical methods, since PECD is already allowed in the pure electric dipole approximation, in contrast to regular CD.\cite{Powis_Advances}
Thereafter, the first experimental observations of PECD were reported for valence photoionization,\cite{Bowering,Garcia_2003,Lischke} and about two years later, a systematic experimental and theoretical study of PECD in core-level photoionization of gaseous camphor confirmed its existence and several features of its behaviour, including its general manifestation within a few tens of eV of an ionization threshold, where the generated photoelectrons are sensitive to the subtleties of any local chiral potential.\cite{Hergenhahn_2004}
Since then, PECD has been studied in the case of one-photon valence and core-level photoionization of gaseous chiral free molecules\cite{Nahon_2015,Turchini_2017,Hadidi_2018} and up to clusters and nanoparticles.\cite{Nahon_2010,Powis_2014,Hartweg2021} Furthermore, its investigation has broadened to include multi-photon\cite{Baumert1,Lehmann_2013,Kastner_2017,Koch, Janssen_2014} and time-dependent\cite{comby2016trpecd,Beaulieu} ionization processes. 
Using charged particle coincidence experiments, the underlying molecular-frame photoelectron angular distributions (PADs) were also measured.\cite{Tia2017,Fehre}
A profound analysis of the symmetry principles underlying the original PECD mechanism and its variants has appeared recently,\cite{Smirnova} and the mechanism underlying the build-up of the asymmetric emission in one-photon PECD has been investigated from a fundamental viewpoint.\cite{SmirnovaA,SmirnovaB} 

Here, we present experimental results in the framework of single-photon photoionization processes in a liquid. The primary question we aim to answer is whether PECD can be observed from the photoionization of a liquid composed of chiral constituents. 
Since the existence of PECD in the gas phase does not require any local molecular ordering, from symmetry principles, this may well be the case. On the other hand, we are not aware of any experiments trying to directly address this question, although a first PECD valence-shell study on pseudo-amorphous nanoparticles of the amino-acid serine revealed a reduced but yet non-vanishing PECD.\cite{Hartweg2021} 
Some of the authors therefore have constructed a new setup dedicated to PECD studies on a liquid microjet, as described elsewhere.\cite{EASI}
Here, we present a complete feasibility study of actual PECD detection using a nearly-neat liquid microjet of fenchone. 
This work opens up the perspective to study the handedness of chiral molecules in aqueous solution, such as amino acids, their building blocks,\cite{Nolting} or sugars.\cite{Mudryk}

Fenchone (C$_{10}$H$_{16}$O, 1,3,3-Trimethylbicyclo[2.2.1]heptan-2-one) is a chiral bicyclic mono-terpenoid built from a six-membered ring with a single-carbon bridge connecting C1 and C4 and featuring several methyl ligands and a carbonyl (C=O) group adjacent to one of the asymmetrically substituted chiral centres. 
A structural diagram is shown below in Fig.~\ref{fgr:overview}. 
The (1R,4S)-fenchone enantiomer naturally occurs in fennel. Importantly, the {\bf C}=O carbonyl carbon has a 1s core binding energy shift that allows it to be spectroscopically distinguished from the remaining carbon atoms in a  core-level photoelectron spectrum (PES).\cite{EASI} Moreover, it has been shown to exhibit a sizeable PECD effect in the gas phase.\cite{Ulrich,EASI}
Follow-on studies on this molecule also examined PECD effects in its valence PES,\cite{Fenchone1,Fenchone2} and subsequently targeted multi-photon PECD processes\cite{Baumert1,Kastner_2017,Kastner_2020, Singh_2021} and complex electronic-structure dynamics using ultrafast laser pulses.\cite{comby2016trpecd,Beaulieu_2018} The choice of fenchone for these prototypical studies has been partially motivated by the relative rigidity of its geometric structure, making conformational isomerism a lesser complication in the interpretation of any associated results, in comparison to those from other similarly-sized chiral systems. 
In this work, we will address single-photon C~1s core-level PES of fenchone in its native, liquid form, as this presents a clear-cut case for the demonstration of liquid-phase PECD.
A small subset of the data from this project was already used for illustrative purposes in an apparatus paper that some of the authors have recently published.\cite{EASI}

\section{Experimental}
Photoionization experiments on a liquid microjet of fenchone were performed using circularly polarized synchrotron radiation from an undulator, and a hemispherical electron analyzer arranged in the backward-scattering plane. 
Data were acquired over two measurement campaigns with a setup described recently.\cite{EASI} 
Details of the experimental setup are as follows:
\subsection{Synchrotron Radiation}
The experiments used synchrotron radiation in the soft X-ray range provided by the P04 beamline of the PETRA~III storage ring at DESY, Hamburg (Germany). This beamline is equipped with an APPLE-II-type undulator\cite{Sasaki} allowing experiments with high-purity CPL.\cite{P04} 
A VLS (variable line spacing) monochromator's planar grating of 400~l/mm spacing and 15~nm groove depth (campaign 1) or 1200~l/mm spacing and 9~nm groove depth (campaign 2) was used with typical exit-slit settings of 100-120~µm, yielding an energy resolution of approximately 90~meV (400~l/mm) or 30~meV (1200~l/mm) at photon energies slightly above the carbon K-edge. 
The minimum spot size of the beamline (using smaller exit-slit openings than 30~µm) has been measured as (h x v) 10 x 10~µm$^2$ in the nominal focus position of the optics.\cite{Focusfinder,Bagschik} 
For our experiment, due to spatial constraints, the interaction region had to be placed approximately 220~mm downstream of that position, the vertical focus however was shifted accordingly using a pair of Kirkpatrick-Baez mirrors.\cite{P04} 
We correspondingly estimate the beam spot size at the point of interaction to be (h $\times$ v) 180 $\times$ 35~µm$^2$.

The photon-energy scale of the monochromator was calibrated by a standard procedure that optimizes the pitch angle for specular reflection. 
We estimate the residual error after this procedure as $\pm 0.2$~eV in the photon-energy range used in this work.

Circularly polarized radiation of either helicity was produced by shifting the magnet blocks of the APPLE-II undulator accordingly.
In a separate experiment, the polarization of the ensuing radiation has been analyzed by measuring the photoelectron angular distributions (PADs) of various gases in the plane perpendicular to the light propagation direction.\cite{Buck,Circularpolarization}
Measurements were performed for both signs of the undulator geometric shift, corresponding to both output radiation helicities.
The shift was varied in small steps in the spectral region of interest, preferentially yielding circularly polarized light. Then the Stokes parameter for circular polarization was calculated as the complement to the Stokes parameters for the residual amount of linear polarization. Experiments were carried out with the 400~l/mm, 15~nm groove depth grating and yielded absolute values for the circular Stokes parameter, $S_3$, larger than 0.98 for photon energies between 550 and 1250~eV, and in an interval of values of the undulator shift about 5~mm wide. The photon energies of interest in this work are somewhat lower, however, namely in the vicinity of the carbon K-ionization edges of fenchone, slightly below 300~eV. 
In this energy region, some degradation of the purity of circular polarization has been observed in experiments on another APPLE-II undulator beamline, and was attributed to carbon contamination of the optical elements.\cite{Harding_2005} 
In any case, on that occasion $|S_3|$ was still found to be $> 0.92$.
Despite the lack of direct measurements for our setup, we consider it fair to assume similar or lesser circular polarization degradation here. Correspondingly, no further normalization of the measured PECD magnitude has been applied. 

By carrying out a PECD measurement on the fenchone gaseous phase evaporating from the liquid jet, we established a correspondence of the geometric shift with negative sign to left-handed circularly polarized light ({\it l}-CPL), according to the `optical' convention.\cite{Born,Ulrich,EASI}
\subsection{Liquid Microjet}
Both enantiomers of fenchone were obtained commercially (Sigma-Aldrich, purity $\ge 98$\%) and were used without further purification.
A microjet was produced by pushing the liquid through a glass capillary nozzle with an inner diameter of 28~µm by a commercial HPLC pump (Shimadzu LC-20AD).
A flow rate of 0.6-0.8~ml/min at pressures of 11-14~bar was typically used.
The sample was made conductive by addition of 75~mM tetrabutyl-ammonium nitrate salt (TBAN), to prevent charging by the photoionization process.
Our liquid-jet holder features a cooling jacket that was stabilized to 10\,$^\circ$C. 
Since it, however, does not extend up to the nozzle tip, a slightly higher temperature of the injected liquid cannot be ruled out.
The liquid stream was directed horizontally, perpendicular to the light propagation axis.
After passing the interaction region, the jet was collected on a cold trap cooled by liquid nitrogen, thus maintaining the interaction chamber pressure below 10$^{-3}$~mbar.
A bias voltage could be applied to the liquid microjet via a gold wire brought in contact with the liquid approximately 550~mm upstream of the expansion nozzle;\cite{EASI,Thuermer2021}
this wire was connected to the chamber ground potential, unless otherwise stated.
Comparison measurements were performed using the same equipment to produce a jet of high-purity liquid water, made conductive by the addition of NaCl to 50~mM concentration.
\subsection{Electron Detection}
Photoelectrons produced from the liquid fenchone jet by circularly polarized synchrotron radiation  were detected in a backward-scattering geometry under an angle of 130$^\circ$ with respect to the light propagation direction. 
Electrons were detected by a near-ambient-pressure hemispherical electron analyzer (HEA, Scienta-Omicron HiPP-3) with a lens mode adapted to specifically enable the measurement of electrons at low kinetic energies (KEs, below 30~eV).
For the same purpose, $\mu$-metal shielding was added to the interaction chamber housing the liquid jet. 
Electrons passed a first skimmer into the HEA with an opening of 800~$\mu$m diameter and set to the ground potential of the setup, and were accelerated immediately thereafter to diminish scattering losses at the elevated background pressure produced by evaporation from the liquid jet.
The distance of the liquid jet to this opening was approximately equal to the skimmer aperture diameter.
Under these conditions, photoelectrons from a liquid jet can be observed down to very low KEs, though they appear atop of an intense background of low-energy electrons produced by scattering of photoelectrons created inside the bulk liquid.\cite{Malerz}
This point will be further discussed in detail below.

The slit restricting the entrance into the hemispheres was set to 800~$\mu$m, adapted to the size of the skimmer opening. 
Electron spectra were measured with a pass energy of 20~eV, and electrons were detected by a stack of two microchannel plates and a fluorescence screen, read out by a CCD camera.
The so-called ADC (analog-to-digital conversion) mode of the control software was used, in which the gray-scale camera image is interpreted to yield the underlying electron detection rates.

Spectra were acquired in swept mode. 
In order to minimize loss of acquisition time by shifting the undulator structure and switching the X-ray beam helicity, spectral sweeps were typically repeated ten to thirty times for each photon helicity, and several pairs of spectra were acquired for both helicities at each photon energy.
A set of individual sweeps that were averaged to produce a spectrum are shown in Fig.~S1 of the ESI.\dag\ 
Some amount of sweep-to-sweep variation is seen, concerning both intensities and peak energies.
The typical extent and time-scale of intensity fluctuations is further illustrated by Fig.s~S2 and S3 of the ESI.\dag\ 
While the exact origin of these effects is still under investigation, the occurrence of small variations of the jet position (much smaller than the focus size, that is on a length scale of one-two µm) likely contributes significantly to these observations.
Before analysing the intensity difference between {\it l,r}-CPL, the raw data were checked for sweep-to-sweep variations of intensity or KE, and sweeps identified as clear outliers were removed. 
Intensities were always determined from sweep-averaged spectra, to make up for the fact that a different amount of sweeps may pass the quality criterion for {\it l}- vs.\ {\it r}-CPL.
Between 3-30\% of sweeps were dropped.
In some cases, small KE drifts over the course of data acquisition (50~meV or less) were numerically corrected.
Methods for peak-area determination and peak-to-background separation were an essential part of the data analysis and will be detailed below.
\section{Results}
\begin{figure}[ht]
\centering
  \includegraphics[width=8.3cm]{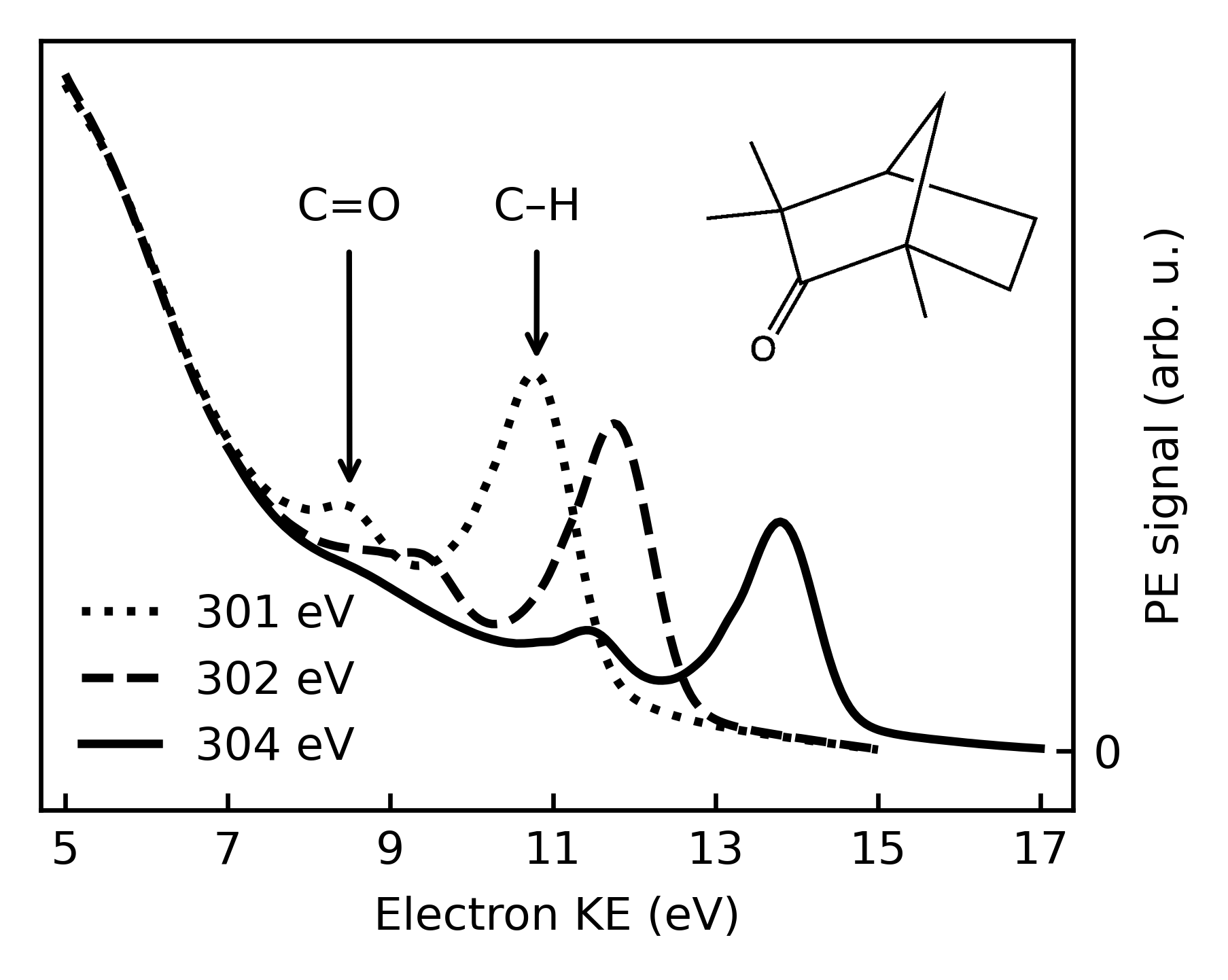}
  \caption{Electron spectra recorded after photoionization of (1S,4R)-(+)-fenchone with {\it l}-CPL at photon energies of 301 eV, 302 eV, and 304 eV; dotted, dashed, and solid lines, respectively.
  Spectra averaged over a number of sweeps performed at the respective photon energy are shown; no further normalization has been performed.
  See text for details.}
  \label{fgr:overview}
\end{figure}
\subsection{The C~1s photoelectron spectrum of liquid fenchone}
Figure~\ref{fgr:overview} shows typical C~1s photoemission spectra from liquid (1S,4R)-fenchone, measured at 301 eV, 302 eV, and 304 eV photon energies with {\it l}-CPL.
Two features due to C~1s photoionization can be readily identified and strongly resemble earlier results for gaseous fenchone.\cite{Ulrich}
The less intense peak at lower KE (higher binding energy) is correlated with ionization of the single carbon from the C=O double-bond carbonyl group, while the larger peak at higher energy arises from the cumulative ionization of the remaining nine carbon atoms at the primary, secondary, and tertiary sites.
This leads to rather similar C~1s binding energies, which cannot be spectroscopically separated.

No discernible features can be attributed to gas-phase contributions to the spectrum.
This is unusual compared to photoelectron spectra of other substances probed in liquid microjet experiments, most notably water,\cite{Winter_2007} but also, {\it e.g.}, methanol,\cite{Saak} acetic-acid solutions\cite{Hemminger}, and liquid ammonia\cite{Buttersack}.
In principle, two explanations are conceivable: The gas-phase contributions are too low in intensity to become apparent, or they overlap---in this case---with the features stemming from the liquid phase.
The vapour pressure of fenchone in the temperature range relevant for this experiment is 0.33 mbar at 10\,$^\circ$C, more than a factor of ten lower than that of liquid water.\cite{Stejfa_2014}
Typical gas-phase contributions in O~1s spectra of liquid water with the EASI setup at beamline P04 amount to 5-20\% of the signal in the O~1s liquid core level peak, depending on the conditions. Hence, a small gas-phase contribution to the fenchone spectra can be expected.
In our previous work, we deduced an upper limit for the gas-phase contribution of 14\%, based on spectra recorded with a small negative bias applied to the jet in order to separate the gas- and liquid-phase features.\cite{EASI}
From the same analysis, we concluded that gas- and liquid-phase C~1s features indeed energetically overlap in the current case.
This is a rather non-trivial result, as even in the valence spectrum of liquid fenchone (unpublished data from our own work), or of other non-polar, liquid solvents,\cite{Chergui,Schewe} ionization energies are typically lower in the liquid in comparison to the gas phase.
In a crude manner, the gas-liquid shift was rationalized by considering the Born free energy of solvation of a positive charge (the vacancy created by photoionization) in the bulk liquid, described by its polarizability, $\varepsilon$, at optical frequencies.\cite{Barth}
The quantity $\varepsilon$, taken as the square of the refractive index, does not differ qualitatively between fenchone and liquid water.\cite{CRC}
Therefore, we suggest that the small or vanishing gas-liquid shift for the inner-shell levels of fenchone is coincidental; it may result from a cancellation of various factors, {\it e.g}.\ electronic charge redistribution following ionization versus electronic structure changes due to nuclear rearrangement.
In this study, we additionally append the previously-determined gas-phase binding energies of 292.6~eV (C=O site) and 290.3~eV (CH site) to the analogous liquid phase peaks.\cite{Ulrich}

In addition to the C~1s main lines, an unstructured background of low-KE electrons can be seen (low KE tail, LET).
This phenomenon is well known from photoemission studies on bulk solid samples\cite{Hufner} and has recently been described in detail for a liquid water jet by some of the authors.\cite{Malerz}
Briefly, in our study on aqueous solutions, an intense LET was found, atop of which no discernible structures could be resolved in electron spectra below kinetic energies of approximately 10~eV.
This is a general result, valid not only for emission out of water's orbitals, but also for features resulting, {\it e.g.}, from electronic levels of solutes.\cite{Malerz}
While its exact nature is not fully understood at this moment, it is attributed to a strong increase of the importance of quasi-elastic, {\it e.g. }vibrational scattering channels, particularly at electron kinetic energies for which electron-impact ionization channels are closed.
Adding to that is an influence of excitation into neutral resonant states lying above the nominal ionization potential.\cite{Malerz}

Strictly speaking, the nature of the LET and the phenomenon of diminishing peak intensities may very well be of a different nature in fenchone, {\it e.g.}, less or more intense and with a different energetic threshold, since in liquids little is known about the LET dependence on the ionized substance. 
Fig.~\ref{fgr:overview} suggests that peak features are observable with acceptable spectral distortion down to KEs of 8 eV in liquid fenchone, which is similar or slightly lower compared to water.
This result is of great importance for our work, as in gas-phase studies it has been learned that PECD only leads to significant asymmetries in the threshold region, {\it i.e.}, at photoelectron kinetic energies below 20 eV. 
Notably, a comparison of the results on low-KE electron emission from liquid water in Ref.~\citenum{Malerz}, giving a lower KE bound at which liquid-phase photoemission peaks from aqueous solutions are discernible, with the gas-phase data by Ulrich {\it et al.,}\cite{Ulrich} giving an upper KE bound at which PECD is still sizeable, suggests that the energy window shown in Fig.~\ref{fgr:overview} spans a range offering good prospects for the identification of PECD in a liquid.
\subsection{Observed dichroism in the angle-resolved spectra}
In order to demonstrate the functionality of our apparatus the PECD of gas-phase fenchone, as sampled by lowering the liquid jet out of the synchrotron-radiation focus, was recorded and the literature results of Ulrich {\it et al.}\cite{Ulrich} were successfully reproduced with an improved energy resolution and a shorter acquisition time.
These tests are described in our recent apparatus description and characterization paper.\cite{EASI}
\begin{figure}[h!tb]
\centering
  \includegraphics[width=8.3cm]{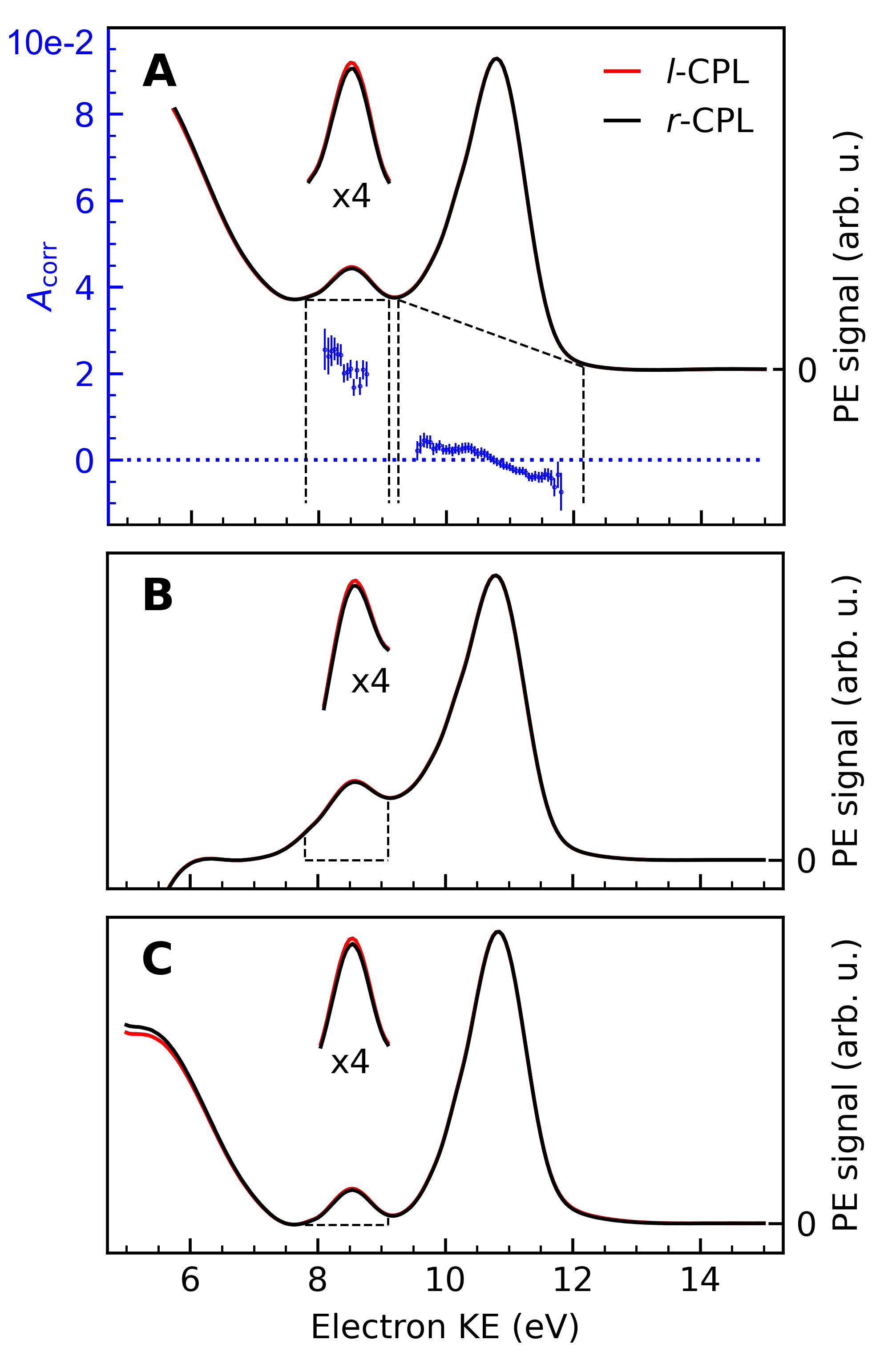}
  \caption{Background-corrected photoemission spectra of (1S,4R)-(+)-fenchone measured at 301~eV with {\it l}- and {\it r}-CPL.
  Panels (A) to (C) show the results of various models by which to subtract the background contribution from a pair of spectra; spectra in Fig.~\protect\ref{fgr:overview} were instead displayed as measured for {\it l}-CPL.
  Background contributions were calculated by (A) fitting an exponential function to the high kinetic end of the spectrum, and subtracting an  additional linear background (`roi'-approach); or (B) by fitting a linear combination of an exponential and a linear function to the high- and low-KE ends of the spectra (`exp'-approach). For (C), an exponential function has been fitted and subtracted from the raw data. Then, the `total background' function\cite{Li} is applied to the remaining spectrum (`sum'-approach). Blue points in panel (A) indicate the experimental asymmetries, $A_\text{corr}(130^{\circ})$  (plotted versus the left y-axis) for the peaks originating from {\bf C}=O and C-H K-edge photoionization, obtained as the difference divided by two times the mean of {\it r}- and {\it l}-CPL (eqn~(\protect\ref{eq:Acorr})), together with associated error bars. 
  The areas marked with dashed lines in panels (B) and (C) indicate the width of the {\bf C}=O peak, which is used for the asymmetry calculations. 
  Expanded ($\times4$) views of these peaks are drawn above the full spectra to get a clearer view on the magnitudes of the {\it l-r} asymmetry.
  \label{fgr:backgrounds}}
\end{figure}

We now turn to an analysis of the differences in photoemission spectra recorded with different helicities of the ionizing photons.
Conceptually, we will distill an intensity asymmetry due to PECD from pairs of spectra recorded in the energy range shown in Fig.~\ref{fgr:overview} by taking the following three steps:
\begin{enumerate}
    \item peak-to-background separation and background subtraction,
    \item calculation of the asymmetry from a pair of spectra at equal KE, and
    \item correction of this raw value for any apparatus asymmetry.
\end{enumerate}
As the first step, quantifying the amount of background present underneath the two C~1s peaks turned out to be the most problematic as obviously the C~1s signal is outweighed by the background contribution.
Representing it by an analytic procedure recommended for UPS data\cite{Li} did not yield a satisfactory representation.
We therefore tested several {\it ad hoc} approaches to background subtraction, and compare them in detail below.

An exemplary background-corrected pair of sweep-averaged spectra of (1S,4R)-fenchone measured at 301~eV with {\it l}- and {\it r}-CPL is shown in Fig.~\ref{fgr:backgrounds}, for all three background models used.
Before we detail the various background-subtraction methods further, we would like to discuss dichroic properties of these two spectra and our approach to apparatus asymmetry correction.

In panel (A) of Fig.~\ref{fgr:backgrounds}, we show a pair of spectra, averaged over two equally long sets of sweeps for each helicity after deletion of outlier traces and correction of (small) energy drifts.
A visible apparatus asymmetry due to a small mismatch in photon intensities produced by the undulator in its two opposite settings has been corrected for in the figure as detailed below.
The deviation of the intensity ratio from unity by this effect is practically invariant over the narrow photon-energy interval targeted in this paper, and is constant over a measurement campaign.
Uncorrected spectra are shown in Fig.~S4 of the ESI; the intensity mismatch can also be seen in the general trend of the per-sweep total intensities in Fig.~S2 of the ESI.\dag\ 
In order to correct for this apparatus-induced effect, we have used the helicity-dependent intensity of the C-H peaks in the spectra as an internal light-intensity monitor.
This follows a practice from gas-phase studies of PECD in several terpenoids featuring a single C=O double bond, where it was noted that an asymmetry associated with the sum of overlapping hydrocarbon site signals in the more intense C-H peak can reasonably be assumed to cancel out.\cite{Hergenhahn_2004}
It was therefore postulated that the asymmetry of the latter peak vanishes exactly, and the C=O asymmetry was correspondingly measured relative to it.\cite{Hergenhahn_2004}
While this started out as an {\it ad hoc} assumption, this tenet was experimentally verified after work on the data-acquisition methodology allowed the measurement of gas-phase PECD free from a baseline error.\cite{Harding_2005,Ulrich}
We have, therefore, determined the asymmetry of the C-H peak as explained below. For the purpose of illustration, we have used this information to normalize the pair of traces shown in Fig.~\ref{fgr:backgrounds} such that they  correspond to the outcome of a measurement that is free from apparatus-induced asymmetry.
Here and in the following we use the ratio:
\begin{equation}
A = \frac{L-R}{L+R} = \frac{r-1}{r+1}\quad\hbox{with}\quad r\equiv \frac{L}{R}
\end{equation}
to calculate the asymmetry, $A$, from the intensities $L$ and $R$ recorded with {\it l}-CPL and {\it r}-CPL, respectively.
If we include a correction for the apparatus asymmetry, the corrected asymmetry $A_\text{corr}$, determined from a measured intensity ratio $r'$ and a correction factor $\gamma$ is then:
\begin{equation}
A_\text{corr} = \frac{r'\gamma-1}{r'\gamma+1},
\label{eq:Acorr}
\end{equation}
where $\gamma$ can be determined from the measured asymmetry $A_0'$ of an isotropically emitted line (intensities $L_0', R_0'$) by:
\begin{equation}
\gamma = \frac{1-A_0'}{1+A_0'} = \frac{R_0'}{L_0'}.
\end{equation}

In that terminology, panel (A) of Fig.~\ref{fgr:backgrounds} shows the traces $L\gamma$ and $R$.
An exponential function fitted only to the part of the spectrum at higher KE than the C-H main line has also been subtracted from the data.

An intensity difference in the traces corrected for apparatus asymmetry, shown in Fig.~\ref{fgr:backgrounds}A, can be seen in the region of the C=O C~1s line at a KE of $\sim$8.5~eV.
In order to exclusively analyse the intensity that can be attributed to C~1s photoemission, we have subtracted a further, linear background, as indicated by the dashed lines.
Within the main C~1s peaks, we have then calculated the corrected asymmetry $A_{\text{corr},i}$ for every KE channel $i$. 
The resulting values are shown in Fig.~\ref{fgr:backgrounds}A, plotted against the left-hand axis in a blue color.
The error bars were derived as follows: We arranged all sweeps made with {\it l}- and with {\it r}-CPL into pairs.
Labelling the pairs with the index $k$, we then calculated the distribution $A_{\text{corr},i}^k$, and give the standard deviation of its mean as an error to the data point $A_{\text{corr},i}$.
More details of the data processing steps are provided in ESI\dag\ Section 1.3 and Supplementary Fig.s~S4,S5.

Trivially, channel-wise asymmetries $A_i$ are equal or very near to zero in the range of the major C-H peak (showing the consistency of the baseline correction). 
Whether the rising trend of the asymmetry data pointing from slightly negative to slightly positive values in going towards smaller KEs is significant cannot be definitively ascertained at this moment.
A word of caution is needed about its interpretation, as minute differences in the peak profile as a function of KE may occur {\it e.g.\ }due to small pointing differences for the left-handed versus right-handed undulator radiation, and can readily produce the apparent trend.
On the contrary, asymmetry values for the C=O peak clearly show an asymmetry which is significantly different from zero.
Still, with the current data and uncertainty limits, we refrain from (over-)interpreting the trend of the C=O asymmetry data across the low KE C 1s peak.
\subsection{Analysis of the C~1s peak areas}
No clear-cut approach to peak-to-background separation is applicable to our spectra (see Fig.~\ref{fgr:overview}), to the best of our knowledge.
As this point is nevertheless essential, we used different methods in parallel and will compare their results in subsection \ref{ss:avgresults}.
Panels (A)-(C) in Fig.~\ref{fgr:backgrounds} serve to illustrate these methods.

As explained above, an exponential background was subtracted in Fig.~\ref{fgr:backgrounds}A. 
Subsequently, peak areas were determined as integrated counts between the range marked with the vertical dashed lines, minus a linear background as indicated by the approximately horizontal dashed lines.
Using the term `region of interest' for these ranges, we term this the `roi' method.

The spectra in Panel (B) were constructed by selectively fitting a linear combination of an exponential and a linear function to the data points containing the LET contribution only, specifically at the low- and high-KE ends of the spectra. After subtraction of the estimated background, the spectra were normalized to the C-H peak maximum. Asymmetries have then been calculated from the integrated PE intensities in a 1.4~eV energy range around the C=O peak, as indicated by the areas enclosed by the dashed lines in Panel (B) of Fig.~\ref{fgr:backgrounds}.
We use the label `exp' for this method.

The approach adopted to produce Panel (C) follows a similar procedure to that used to produce Panel (B), with the exception that the background was constructed by first fitting an exponential function to the high-KE side of the spectra and then applying the `total background' function\cite{Li} (also known as non-iterative Shirley method \cite{Vegh2006}) in order to estimate the LET background. This procedure iterates from the high- to the low-KE end of the spectra while aggregating (`summing') spectral intensity and is thus referred to as `sum'-approach.

More detail on the various background-signal separation methods is provided in the ESI.\dag\ 
\subsection{Parametrization of the measured results}
In order to connect our results to other experiments and to calculations, it is important to resort to parametrized forms of the differential photoionization cross section, which is the quantity measured here.
Building on the earlier work of Ritchie,\cite{Ritchie1,Ritchie2} Ivan Powis showed that within the electric-dipole approximation, for chiral molecules the differential photoionization cross-section can be cast in the following form:\cite{Powis_2000,Powis_Advances}
\begin{equation}
I^p(\theta) = \frac{\sigma}{4\pi} \left[1 + b_1^p\cos(\theta) + b_2^pP_2(\theta)\right].
\label{eq:ang}
\end{equation}
Here, the intensity has been written as a function of the angle $\theta$ measured from the photon propagation vector to the electron emission vector, for the left-handed circular ($p=1$) or right-handed circular ($p=-1$) pure polarization states. 
$P_2$ denotes the second Legendre polynomial.
The second-order coefficient $b_2$ turns out to be independent of the handedness of circular polarization, and can be expressed via the more familiar $\beta$-parameter as $b_2^{+1}=b_2^{-1} = -\beta/2$.
A similar equation can be written for linearly polarized ($p=0$) light, with the understanding that the angular coordinate in this case refers to the major axis of the polarization ellipse.
In the latter case, the first-order coefficients vanish ($b_1^0 = 0$) and the second-order coefficient becomes $b_2^0 = \beta$.
Higher-order and magnetic terms in the interaction of the radiation field with the molecule or liquid have been presented,\cite{Ritchie2} but based on experimental results, they seem to have been unimportant in earlier gas-phase work.\cite{Hergenhahn_2004}
The first-order coefficient $b_1$ vanishes for non-chiral molecules, and undergoes a sign change upon changing the chiral handedness of the molecule or swapping the light helicity; it is therefore the chiroptical parameter defining the PECD-induced asymmetry. Correspondingly, we can identify the corrected asymmetry (\ref{eq:Acorr}) as follows:
\begin{equation}
A_\text{corr} = 
\frac{I^{+1}(\theta) - I^{-1}(\theta)}{I^{+1}(\theta) + I^{-1}(\theta)}
= \frac{b_1^{+1}\cos\theta}{1-(\beta/2)\,P_2(\cos\theta)}.
\label{eq:Apara}
\end{equation}
Here, the symmetry relation $b_1^{+1} = -b_1^{-1}$ has been used.\cite{Ritchie1,Powis_2000}
It is interesting that, in the general case, $A_\text{corr}$ depends on both the chiral parameter $b_1$ and the conventional angular distribution.
In earlier gas-phase PECD experiments based on measurements performed at a single electron collection angle,\cite{Hergenhahn_2004} or a pair of angles mirrored in the dipole plane,\cite{Turchini_2004} the so-called magic angle-geometry of $\theta=54.7^\circ$ was used. This geometry was adopted to ensure that the denominator in eqn\,(\ref{eq:Apara}) becomes identical to unity and the dependence on $\beta$ ceases. Another option is to use imaging techniques collecting electrons over the full 4$\pi$ sr emission solid angle, directly allowing the cosine dependence of the asymmetry to be extracted.\cite{Garcia_2003,Nahon_2006}

In our case, a rigorous determination of $b_1$ from our experiment would require a separate measurement of $\beta$, which however was outside the scope of this work.
In the following, we will therefore estimate the potential influence of the deviation of our setup from the magic-angle geometry.
Similarly, an estimate of the potential influence of non-complete circular polarization is in order.
Inserting the limiting values of $\beta$ ($-$1 and +2) into $[1-(\beta/2)\,P_2(\cos\theta)]^{-1}$, we find that this factor may range from 0.94 to 1.14 for our $\theta = 130^\circ$ detection geometry.
In the few works on the angular distribution parameter in photoemission from liquids, however, a trend towards small absolute $\beta$ values has been found  at low kinetic energies.\cite{Thuermer_O1s}
Given the results of that study on the O~1s orbital of water, a $\beta \sim 0.5$ might be a plausible but conservative estimate for our case of C~1s emission, which would lead to a factor of 1.03, resulting from the denominator in eqn\,(\ref{eq:Apara}).

Further, consideration of polarization impurities requires a look at the full angular-distribution function, which can be written as:
\begin{align}
&I(S,\theta,\phi) =\nonumber\\ 
&\frac{\sigma}{4\pi} \left[1 - S_3\,b_1^{+1}\cos\theta - \frac{\beta}{2}\,\bigl( P_2(\cos\theta) - \frac{3}{2}(S_1\cos 2\phi + S_2\sin 2\phi)\sin^2\theta \bigr)\right],
\label{eq:fullpad}
\end{align}
with the understanding that the first $\theta$-dependent term is only present for chiral molecules.\cite{schmidtbook,Powis_2000}
The polarization state of the radiation is now represented by the three-component Stokes vector $S$, with $S_1$ and $S_2$ representing linear polarization measured with horizontal and vertical, or 45$^\circ$ and 135$^\circ$ polarizers, and $S_3$ defining the degree and type of circular polarization.
The angle $\phi$ is measured from the horizontal axis in the dipole plane to the electron spectrometer, and amounts to 90$^\circ$ in our experiment.
As explained in the experimental section, polarimetry results in the photon-energy range of interest are not available for the P04 beamline yet.
However, the degree of linear polarization and the direction of the polarization ellipse were measured between 550 and 1250~eV, as a function of the undulator shift.\cite{Circularpolarization}
The complement of the linear polarization degree was attributed to circular polarization, neglecting the presence of an unpolarized fraction of radiation.
This is supported by full polarimetry results associated with another APPLE-II undulator beamline.\cite{Wang}
In all data sets recorded, when the circular component was maximized, a remaining Stokes parameter of linear polarization with magnitude 0.04 or smaller was found, which almost exclusively had $S_2$ character.
As, from eqn (\ref{eq:fullpad}), $S_2$-dependent terms cannot play a role in our geometry, we will neglect the residual linear components entirely, although this is not fully rigorous.
Including a finite angular acceptance of our electron analyzer (see Ref.\  \citenum{EASI}) in a discussion of eqn (\ref{eq:fullpad}) leads to corrections that vanish, to first order, and are essentially independent of the enantiomer and helicity of the light. 
We therefore deem it safe to assume that the impact of such effects is much smaller than the others we have explicitly considered above

To summarize this discussion, we find that optical polarimetry at the carbon edge, to determine the on-target radiation state, and a photoelectron angular-distribution measurement on liquid fenchone would be desirable for a quantitatively accurate determination of the $b_1$-parameter in our experiment. However, for the moment we will retain the simple relation, $b_1^{+1} = A_\text{corr}/\cos\theta$, and will make an appropriate adjustment to the error bar with respect to the influence of $\beta$ and any residual non-circular soft X-ray beam polarization.

\begin{figure*}
 \centering
 \includegraphics[width=17.1cm]{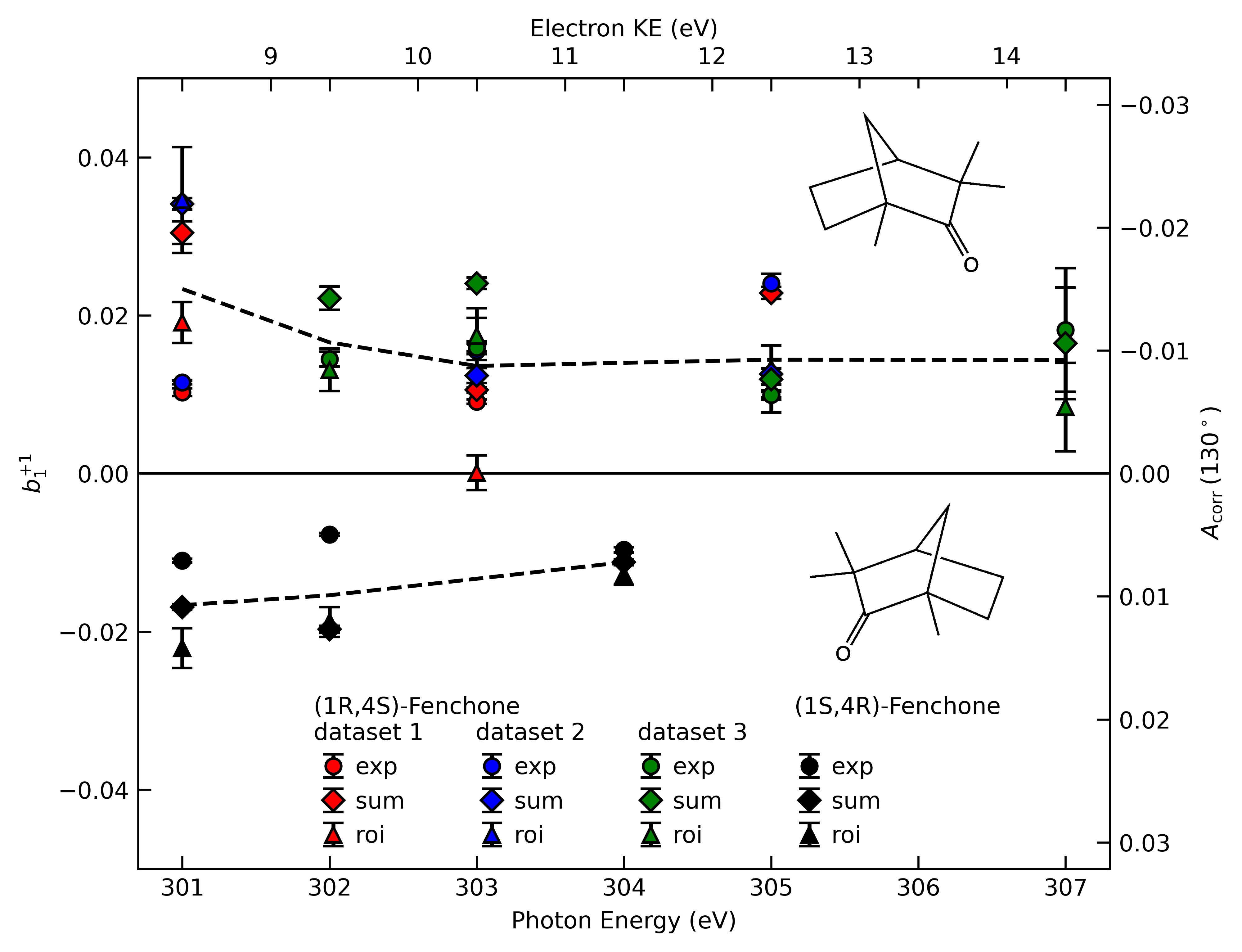}
\caption{The corrected asymmetry, $A_\text{corr}$, and the resulting chiral angular-distribution parameter $b_1^{+1}$ as a function of photon energy.
We use black symbols for the (1S,4R)- and colored symbols for the (1R,4S)-enantiomer.
For the latter, the results from three data sets, acquired in two different measurement campaigns, are shown to indicate the stability of our experiment.
Different approaches to subtract the LET and, possibly, a residual background are differentiated by the symbol shape, with diamonds referring to the `total background' approach (Fig.~\protect\ref{fgr:backgrounds}C, `sum'), circles to the linear-exponential approach (Fig.~\protect\ref{fgr:backgrounds}B, `exp') and triangles to the region-of-interest (`roi') approach (Fig.~\protect\ref{fgr:backgrounds}A).
To guide the eye we indicate the averaged values detailed in Table \protect\ref{tbl:b1+1} by dashed lines.  
Values in the figure are not corrected for any possible gas-phase contributions and angular-anisotropy effects (see Table~\protect\ref{tbl:b1+1}).
\label{fgr:b1+1_plot}}
\end{figure*}
\subsection{Averaged and corrected results
\label{ss:avgresults}}
A compilation of the $b_1^{+1}$ values as obtained from the described analysis procedure is provided in Fig.~\ref{fgr:b1+1_plot}. 
The results in the figure have a rather large spread between different data sets and different analysis methods. 
Nevertheless, for most photon energy values, the chiral asymmetry parameter $b_1^{+1}$ is clearly different from zero, with the $b_1^{+1}$ values having an opposite sign for the two different enantiomers. This expected mirroring of the chiroptical data attests that we are indeed measuring, with a reasonable error bar, an enantio-specific observable.

\begin{table*}[h!tb]
\small
  \caption{
Recommended $b_1^{+1}$ values calculated as the averages of the values shown in Fig.~\ref{fgr:b1+1_plot}.
In round brackets, the standard deviation of all values pertaining to the same enantiomer and photon energy is shown.
The rows labelled `measured' are not corrected for the possible presence of gaseous fenchone nor the $\beta$-dependence of the relationship between measured asymmetry and chiral parameter, $b_1$ (see eqn\,(\ref{eq:Apara})).
In rows labelled `corrected', the expected maximum correction of the $b_1$ values has been applied for both factors.
See the main body of the text for details.}
\label{tbl:b1+1}
  \begin{tabular*}{\textwidth}{@{\extracolsep{\fill}}lcccccc}
    \hline
     & 301 eV & 302 eV & 303 eV & 304 eV & 305 eV  & 307 eV\\
    \hline
    measured\\
    (1R,4S)-fenchone  & 0.023(11) & 0.017(5) & 0.014(7) & - & 0.014(6)  & 0.014(5)\\
    (1S,4R)-fenchone  & -0.017(6) & -0.015(7) & - & -0.011(2) & - & - \\
    \hline
    corrected\\
    (1R,4S)-fenchone  & 0.014(12) & 0.010(5) & 0.010(7) & - & 0.011(6)  & 0.012(6)\\
    (1S,4R)-fenchone  & -0.008(6) & -0.009(7) & - & -0.008(2) & - & - \\
    \hline
  \end{tabular*}
\end{table*}
If we scrutinize the data points in Fig.~\ref{fgr:b1+1_plot}, we find that they neither group by analysis method nor by data set.
We therefore believe that the scatter between points does not result from a systematic effect leading to preferentially higher or lower asymmetry values as a function of time or associated with peak-background separation method.
In order to arrive at consolidated values, we performed a simple average over all data points for the same photon energy and enantiomer. 
The results are compiled in Table~\ref{tbl:b1+1}.
The scatter in our data points, perceived as rather coming from fluctuations in the signal and background of the spectra than from the data treatment, is represented by the standard deviation of the individual data points leading to each table entry.
In the table, we also include two potentially important $b_1^{+1}$ corrections, namely one for the presence of gaseous components in the C~1s spectra and another for the potential influence of a non-zero $\beta$-parameter, which would affect the connection between the measured anisotropy and $b_1$ values (eqn\,(\ref{eq:Apara})).
As explained above, although no visible presence of a gaseous component has been observed in the PECD measurements, a separate experiment with a biased jet, albeit at slightly different conditions, suggested that this might result from an inconvenient overlap of liquid-phase and gas-phase C~1s peak features; these results have been discussed in an earlier paper.\cite{EASI}
Referring to that work, we estimate a gas-phase fraction $g$ between negligible, which is expected from the low vapour pressure of fenchone, and $g=0.14$, which is the finding of the aforementioned biased jet experiment.
As the gas-phase contribution has a $b_1$-parameter of larger magnitude, correction for the gas-phase contribution would reduce the liquid phase parameter $b^p_{1,l}$ according to:
\begin{equation}
    b^p_{1,l} = \frac{b^p_{1,m} - gb^p_{1,g}}{1-g},
\end{equation}
where subscripts $m$ and $g$ designate the measured and gas-phase values of $b_1^p$, the latter being taken from Ref.~\citenum{Ulrich} with interpolation where necessary.
For the correction due to the $\beta$-dependent denominator in eqn~(\ref{eq:Apara}), we expect a value between unity (no correction), for a $\beta = 0$, and multiplication by 0.94, for a $\beta = 1$.
%  1/1.064 = 0.9398
Accordingly, the table contains two lines for each parameter stating the averaged, but uncorrected value, and the values corrected downwards by the factors quantified above, which we believe gives the maximum plausible extent of the gas-phase contribution and $\beta$ parameter effects.

We note two further effects that we cannot quantify at this moment, but could be present to some extent.
The exact enantiomeric excess (e.e.) of the samples supplied was not specified and we were unable to have this independently checked, but previous reports have found commercial samples of (1R,4S)-($-$)-fenchone to have a lower e.e.\ than (1S,4R)-(+)-fenchone samples. 
In principle, the measured PECD asymmetry should scale linearly with e.e.\  values when these are known. 
However, such adjustments are here expected to be within the current error bars, and so have not been applied.
The same applies to a correction for an unpolarized fraction of radiation at our sample, which could be slightly increased at the photon energies used in this experiment because of an influence of carbon contamination on the beamline optics.
If present, both factors would lead to a correction of the values of $b_1$ extracted from the measured asymmetry towards larger absolute values.

\section{Discussion}
It is interesting to discuss the reduction in $b_1^{+1}$ relative to gas-phase experiments. 
In the case of fenchone, the reduction in $b_1^{+1}$ amounts to roughly a factor of five. 
This reduction can be compared to results on the conventional angular distribution, represented by the $\beta$ parameter. 
A few experiments for the $\beta$ parameter of photoemission peaks from liquids are available.\cite{Nishitani,Thuermer_O1s,Lewis_2019,Gozem}
In comparison with gas-phase water, a general reduction of $\beta$ has been observed,\cite{Thuermer_O1s,Gozem} but only the study on the O~1s $\beta$ parameter of water by Thürmer {\it et al.\ }extended down to the KEs of interest here.
For their lowest data point at about 12~eV KE, the measured $\beta$-values are approximately $\beta_g = 0.92$ and $\beta_l = 0.28$, which implies a reduction by a factor of 3.3 (with subscripts $g$ and $l$ designating the gas and liquid phase, respectively).\cite{Thuermer_O1s}
Fully consistent with that, the onset of the reduction in $\beta$ upon aggregation of individual molecules was also observed in an experiment on water clusters.\cite{signorell2017clusters}
A plausible explanation for the reduction in $\beta$ is the elastic or quasi-elastic scattering of photoelectrons in the liquid bulk, before traversing the liquid-vacuum interface. 
Due to the random nature of the associated collisions, this would tend to produce an isotropic angular distribution, and the explanation would equally hold for the reduction in $b_1$. 
It could not be shown in Ref.~\citenum{Thuermer_O1s}, however, that this is the sole explanation for the $\beta$ reduction, due to a lack of accurate knowledge of the elastic and inelastic mean free paths of electrons in water. Note that electron scattering was also pointed out as the main source of PECD reduction (by about a factor of five) between nanoparticles and gas phase serine.\cite{Hartweg2021} This effect may be partly compensated by an increased local order in the nanoparticles or fewer associated conformers in the aggregated state. The former explanation may also be applicable to the case of liquid fenchone. 
Elastic electron scattering on the water or fenchone vapour surrounding the liquid jets may additionally contribute to the more isotropic angular distributions from liquids, as the cross-sections for elastic scattering for low-KE electrons on gas-phase water are considerable.\cite{Matsui2016}
As these cross-sections are also strongly peaked at low scattering angles, this will likely be a smaller effect, though.
A redistribution of intensity from the forward- into the backward-scattering plane, which would be necessary for a reduction of $b_1$, is not fully excluded for a cylindrical jet, but seems relatively implausible.

%The \balance command can be used to balance the columns on the final page if desired. It should be placed anywhere within the first column of the last page.

\balance

\section{Conclusions}
A full report on an experiment to measure PECD from the chiral liquid fenchone has been presented.
We have shown a non-vanishing effect of opposite sign for the two enantiomers, with a convincing mirroring attesting the overall quality of the data. 
Akin to studies on the angular-distribution parameter $\beta$ from liquids, and to PECD from homochiral nanoparticles, a substantial reduction of the chiral parameter, $b_1$, has been found relative to the gas-phase sample.
This can be explained to a large or full degree by elastic scattering of the outgoing photoelectrons inside the liquid.
Our study opens up prospects to investigate the solution-phase chemistry of chiral substances in their native environment. The in vivo study of biomolecules in water with simultaneous site- and chemical-specificity, via an analysis of core-level shifts,\cite{Mudryk} and the chiral handedness, {\it via} PECD measurements, is an especially exciting and important example.

\section*{Data Availability}
Data relevant for this study are available at DOI: 10.5281/zenodo.5996526.

\section*{Author Contributions}
Conceptualization: BW, UH, LN, IP; 
Methodology: SM, IW, ST, BW;
Investigation (including data acquisition): MP, SM, FT, CL, CK, LN, IP, IW, BW;
Data analysis (formal analysis): MP, ST, UH;
Visualization, Writing – original draft: UH, MP;
Writing – review \& editing: All authors;
Supervision: BW;
Funding acquisition: BW, UH, SM, DN.
% We strongly encourage authors to include author contributions and recommend using \href{https://casrai.org/credit/}{CRediT} for standardised contribution descriptions. 

\section*{Conflicts of interest}
There are no conflicts to declare.

\section*{Acknowledgements}
The authors would like to acknowledge help from Sebastian Kray and Stefan Schlichting in setting up the experiment, and from Dana Bloss in data acquisition. 
We would like to thank Andr\'e Knie for his contributions in the early stages of this work.
B.W. acknowledges funding from the European Research Council (ERC) under the European Union's Horizon 2020 research and investigation programme (grant agreement No.\ 883759).
F.T. and B.W. acknowledge support by the MaxWater initiative of the Max-Planck-Gesellschaft. 
S.T. acknowledges support from the JSPS KAKENHI Grant No.\ JP20K15229.
D.M.N., M.P., and C.L. were supported by the Director, Office of Basic Energy Science, Chemical Sciences Division of the U.S. Department of Energy under Contract No. DE-AC02-05CH11231 and by the Alexander von Humboldt Foundation.
We acknowledge DESY (Hamburg, Germany), a member of the Helmholtz Association (HGF), for the provision of experimental facilities. Parts of this research were carried out at PETRA~III and we would like to thank Moritz Hoesch in particular, as well as the whole beamline staff, the PETRA~III chemistry laboratory and crane operators for their assistance in using the P04 soft X-ray beamline. The beamtime that enabled this work was allocated for proposal II-20180012 (LTP).

%%%END OF MAIN TEXT%%%

%If notes are included in your references you can change the title from 'References' to 'Notes and references' using the following command:
%\renewcommand\refname{Notes and references}
\renewcommand\refname{References}

%%%REFERENCES%%%
%\eject   % UHe
\bibliography{main} %You need to replace "rsc" on this line with the name of your .bib file

\end{document}

\pagestyle{fancy}
\thispagestyle{plain}
\fancypagestyle{plain}{
%%%HEADER%%%
\renewcommand{\headrulewidth}{0pt}
}
%%%END OF HEADER%%%

%%%PAGE SETUP - Please do not change any commands within this section%%%
\makeFNbottom
\makeatletter
\renewcommand\LARGE{\@setfontsize\LARGE{15pt}{17}}
\renewcommand\Large{\@setfontsize\Large{12pt}{14}}
\renewcommand\large{\@setfontsize\large{10pt}{12}}
\renewcommand\footnotesize{\@setfontsize\footnotesize{7pt}{10}}
\makeatother

\renewcommand{\thefootnote}{\fnsymbol{footnote}}
\renewcommand\footnoterule{\vspace*{1pt}% 
\color{cream}\hrule width 3.5in height 0.4pt \color{black}\vspace*{5pt}} 
\setcounter{secnumdepth}{5}

\makeatletter 
\renewcommand\@biblabel[1]{#1}
\renewcommand\@makefntext[1]% 
{\noindent\makebox[0pt][r]{\@thefnmark\,}#1}
\makeatother 
\renewcommand{\figurename}{\small{Fig.}~}
\sectionfont{\sffamily\Large}
\subsectionfont{\normalsize}
\subsubsectionfont{\bf}
\setstretch{1.125} %In particular, please do not alter this line.
\setlength{\skip\footins}{0.8cm}
\setlength{\footnotesep}{0.25cm}
\setlength{\jot}{10pt}
\titlespacing*{\section}{0pt}{4pt}{4pt}
\titlespacing*{\subsection}{0pt}{15pt}{1pt}
%%%END OF PAGE SETUP%%%

%%%FOOTER%%%
\fancyfoot{}
\fancyfoot[LO,RE]{\vspace{-7.1pt}\includegraphics[height=9pt]{head_foot/LF}}
\fancyfoot[CO]{\vspace{-7.1pt}\hspace{11.9cm}\includegraphics{head_foot/LF}}
\fancyfoot[CE]{\vspace{-7.2pt}\hspace{-13.2cm}\includegraphics{head_foot/LF}}
\fancyfoot[RO]{\footnotesize{\sffamily{1--\pageref{LastPage} ~\textbar  \hspace{2pt}\thepage}}}
\fancyfoot[LE]{\footnotesize{\sffamily{\thepage~\textbar\hspace{4.65cm} 1--\pageref{LastPage}}}}
\fancyhead{}
\renewcommand{\headrulewidth}{0pt} 
\renewcommand{\footrulewidth}{0pt}
\setlength{\arrayrulewidth}{1pt}
\setlength{\columnsep}{6.5mm}
\setlength\bibsep{1pt}
%%%END OF FOOTER%%%

%%%FIGURE SETUP - please do not change any commands within this section%%%
\makeatletter 
\newlength{\figrulesep} 
\setlength{\figrulesep}{0.5\textfloatsep} 

\newcommand{\topfigrule}{\vspace*{-1pt}% 
\noindent{\color{cream}\rule[-\figrulesep]{\columnwidth}{1.5pt}} }

\newcommand{\botfigrule}{\vspace*{-2pt}% 
\noindent{\color{cream}\rule[\figrulesep]{\columnwidth}{1.5pt}} }

\newcommand{\dblfigrule}{\vspace*{-1pt}% 
\noindent{\color{cream}\rule[-\figrulesep]{\textwidth}{1.5pt}} }

\makeatother
%%%END OF FIGURE SETUP%%%

%%%TITLE, AUTHORS AND ABSTRACT%%%
\vspace{1em}
\sffamily

\begin{tabular}{m{2.5cm} p{13.5cm} }

& \noindent\LARGE{\textbf{Photoelectron circular dichroism in angle-resolved\hfill\break photoemission from liquid fenchone:\hfill\break Electronic Supplementary Information}} \\%Article title goes here instead of the text "This is the title"
\vspace{0.3cm} & \vspace{0.3cm} \\

 & \noindent\large{Marvin Pohl,\textit{$^{a,b,c,\ddag}$} 
 Sebastian Malerz,\textit{$^{a,\ddag}$}
 Florian Trinter,\textit{$^{a,d}$}
 Chin Lee,\textit{$^{b,c}$}
 Claudia Kolbeck,\textit{$^{a,\S}$}
 Iain Wilkinson,\textit{$^{e}$}
 Stephan Thürmer,\textit{$^{f}$}
 Daniel M.\ Neumark,\textit{$^{b,c}$}
 Laurent Nahon,\textit{$^{g}$}
 Ivan Powis,\textit{$^{h}$}
 Gerard Meijer,\textit{$^{a}$}
 Bernd Winter,\textit{$^{a}$}
 Uwe Hergenhahn\textit{$^{a}$}} \\%Author names go here instead of "Full name", etc.
 
\end{tabular}

%%%END OF TITLE, AUTHORS AND ABSTRACT%%%

%%%FONT SETUP - please do not change any commands within this section
\renewcommand*\rmdefault{bch}\normalfont\upshape
\rmfamily
\section*{}
\vspace{1cm}

%%%FOOTNOTES%%%

\footnotetext{\textit{$^{a}$~Molecular Physics, Fritz-Haber-Institut der Max-Planck-Gesellschaft, Faradayweg 4-6, 14195 Berlin, Germany. E-mail: hergenhahn@fhi-berlin.mpg.de}}
\footnotetext{\textit{$^{b}$~Department of Chemistry, University of California, Berkeley, CA 94720, USA}}
\footnotetext{\textit{$^{c}$~Chemical Sciences Division, Lawrence Berkeley National Laboratory, Berkeley, CA 94720, USA}}
\footnotetext{\textit{$^{d}$~Institut für Kernphysik, Goethe-Universität Franfurt am Main, Max-von-Laue-Straße 1, 60438 Frankfurt am Main, Germany}}
\footnotetext{\textit{$^{e}$~Department of Locally-Sensitive \& Time-Resolved Spectroscopy, Helmholtz-Zentrum
Berlin für Materialien und Energie, Hahn-Meitner-Platz 1, 14109 Berlin, Germany}}
\footnotetext{\textit{$^{f}$~Department of Chemistry, Graduate School of Science, Kyoto University,
Kitashirakawa-Oiwakecho, Sakyo-Ku, Kyoto 606-8502, Japan}}
\footnotetext{\textit{$^{g}$~Synchrotron SOLEIL, L’Orme des Merisiers, St. Aubin, BP 48, 91192 Gif sur Yvette, France}}
\footnotetext{\textit{$^{h}$~School of Chemistry, The University of Nottingham, University Park, Nottingham, UK}}

\footnotetext{\ddag~These authors contributed equally to this work.}
\footnotetext{\S~Current address: sonUtec GmbH, Mittlere-Motsch-Straße 26, 96515 Sonneberg, Germany.}

%%%END OF FOOTNOTES%%%

%%%UHe
\renewcommand\figurename{Supplementary Fig.}
\renewcommand\tablename{Supplementary Tab.}
\makeatletter 
\renewcommand{\thefigure}{S\@arabic\c@figure}
\makeatother

%%%MAIN TEXT%%%%
\large
\section{Data analysis and quality}
\subsection{Typical raw data}
\begin{figure}[h!tb]
\centering
  \includegraphics[width=12cm]{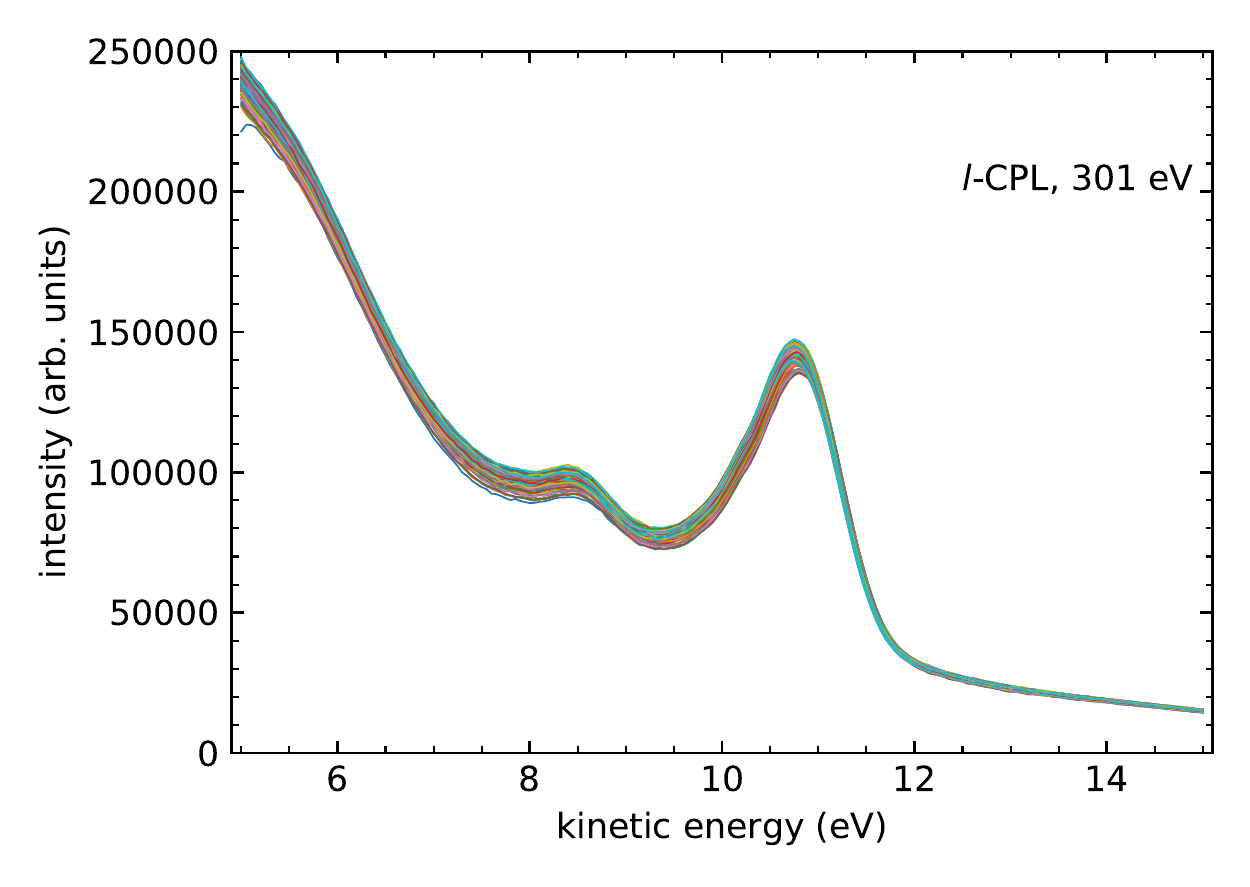}
  \caption{Exemplary, single-sweep photoelectron spectra recorded from liquid (1S,4R)-fenchone, ionized with photons of 301~eV, see the sweep-averaged spectrum in Fig.~1 in the main article. Spectra were acquired in series of thirty sweeps each, with a change of photon helicity after each series. Here, traces from four series, recorded with $l$-CPL are shown. The spectra overlap due to the finite plot resolution.}
  \label{sfig:traces}
\end{figure}
In supplementary Figs.~\ref{sfig:traces} and \ref{sfig:total}, the properties of an exemplary set of liquid fenchone photoemission data are shown.
\subsection{Temporal stability}
\begin{figure}[h!tb]
\centering
  \includegraphics[width=14cm]{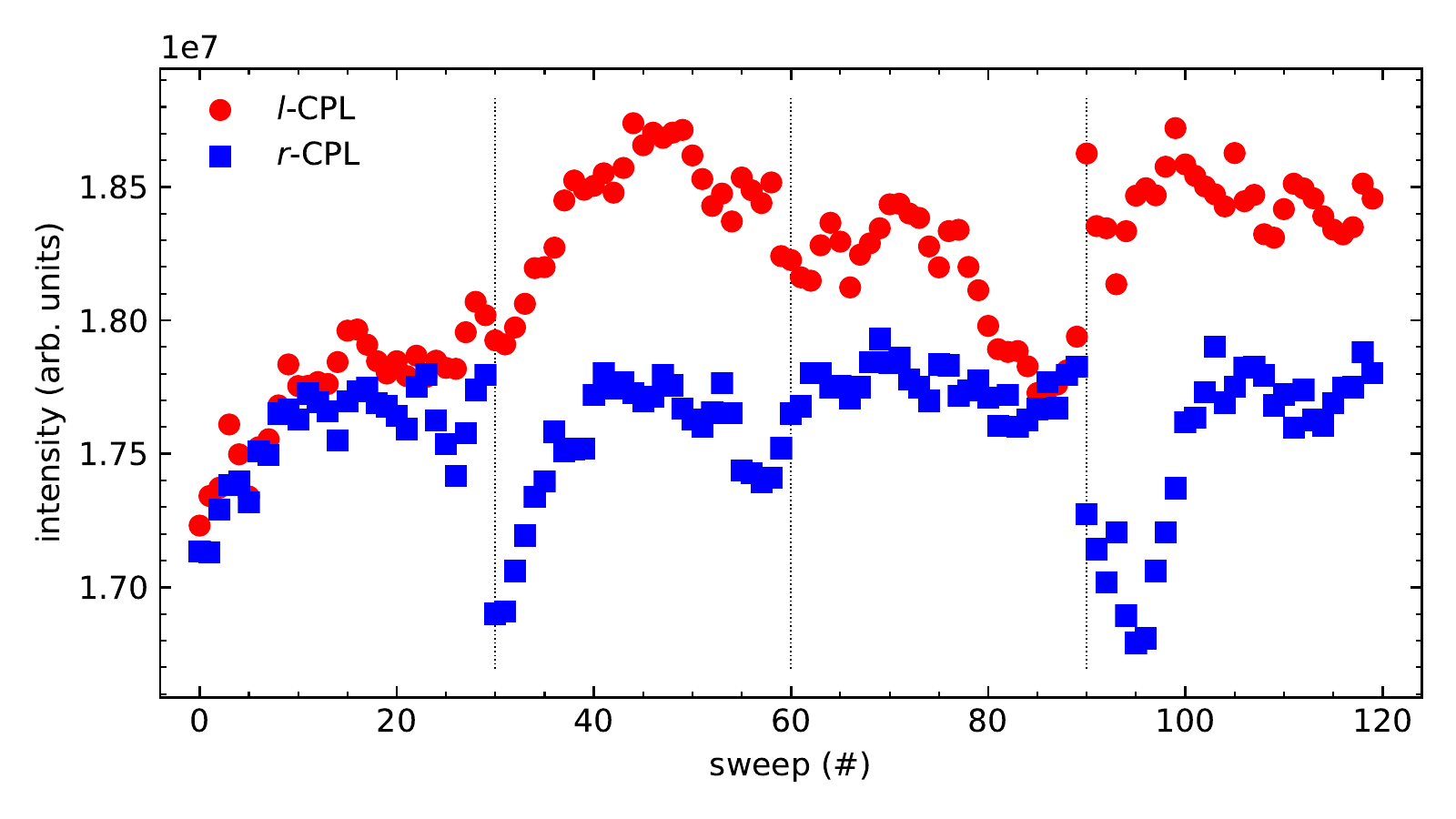}
  \caption{Intensity fluctuations in series of photoelectron spectra from liquid (1S,4R)-fenchone. For each sweep, we show the summed intensity in units yielded by the acquisition software of the electron analyzer. The data shown for $l$-CPL correspond to the sweeps shown in Supplementary Fig.~\protect\ref{sfig:traces}. The vertical dotted lines indicate a shift in helicity, performed after thirty sweeps. The acquisition started by recording sweeps 0-29 with $l$-CPL.}
  \label{sfig:total}
\end{figure}
Small but noticeable fluctuations of the intensity between sweeps are immediately visible in Supplementary Fig.~\ref{sfig:traces}.
The temporal behaviour of the intensity is shown by plotting the integrated counts of every sweep in Supplementary Fig.~\ref{sfig:total}.
\begin{figure}[h!tb]
\centering
  \includegraphics[width=14cm]{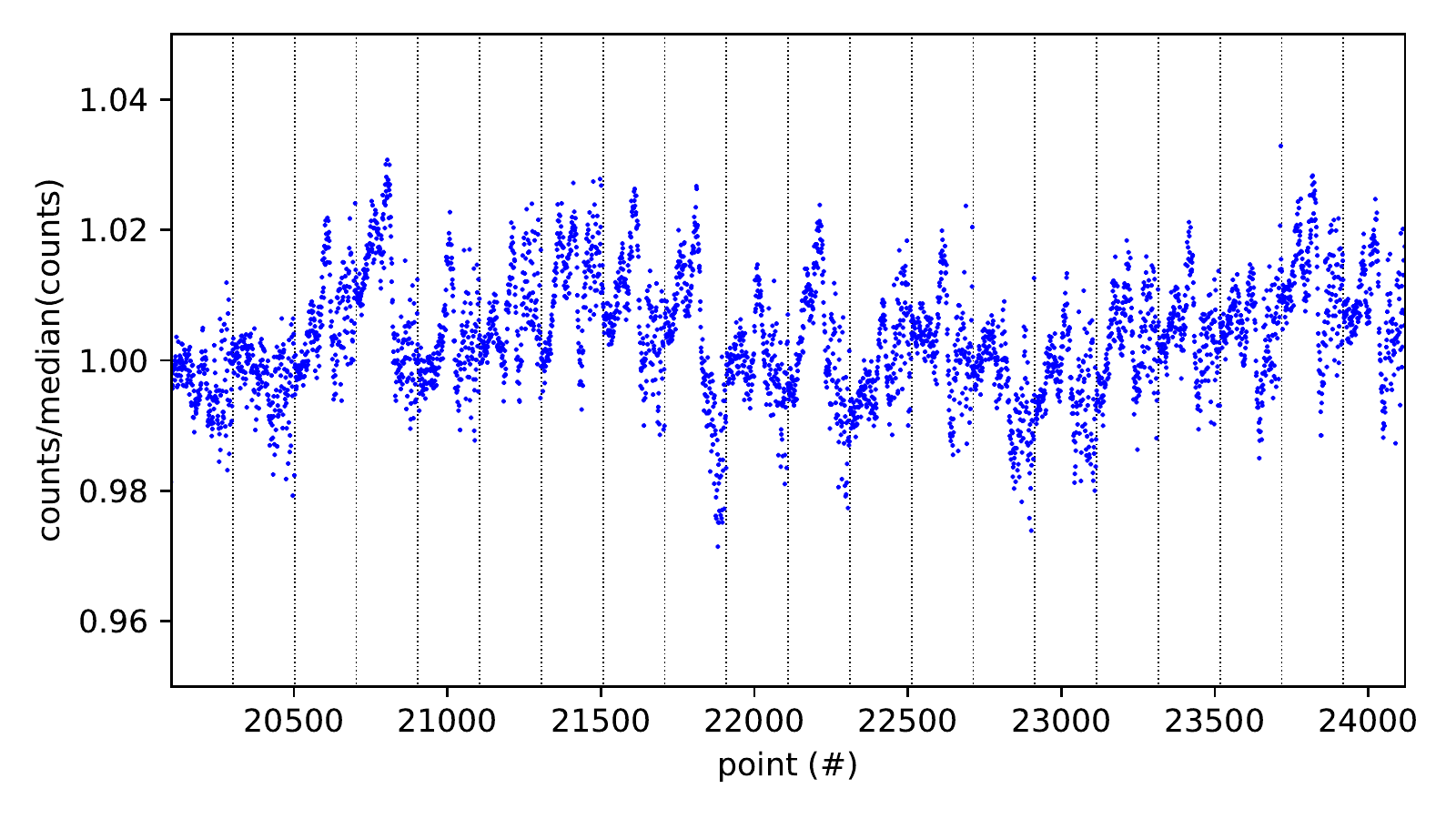}
  \caption{Intensity fluctuations in a series of sweeps to record the photoelectron spectrum from liquid (1S,4R)-fenchone with {\it r}-CPL. For each kinetic energy $i$ in sweep $k$, we plot the counts $a_{i,k}$ divided by the median of $\lbrace a_{i,k};~k = 1,\ldots,120\rbrace$. For plotting, the points have been renumbered consecutively over all sweeps, and the dotted vertical lines show the beginning of a new sweep. A representative interval of the 120 sweeps recorded in total is shown.}
  \label{sfig:ttrace}
\end{figure}
A better view of the fluctuations occurring on a short time scale is offered by plotting the point-to-point variation of the signal intensity shown in Supplementary Fig.~\ref{sfig:ttrace}.
Points were recorded with a dwell time of 0.24~s, the time interval shown in the figure therefore corresponds to an acquisition time of approximately 1350~s, including a time margin allowed for settling of the analyzer voltages as the centrally analysed electron kinetic energy is swept.
\subsection{Peak area integration, `roi'-method}
\begin{figure}[h!tb]
\centering
  \includegraphics[width=12cm]{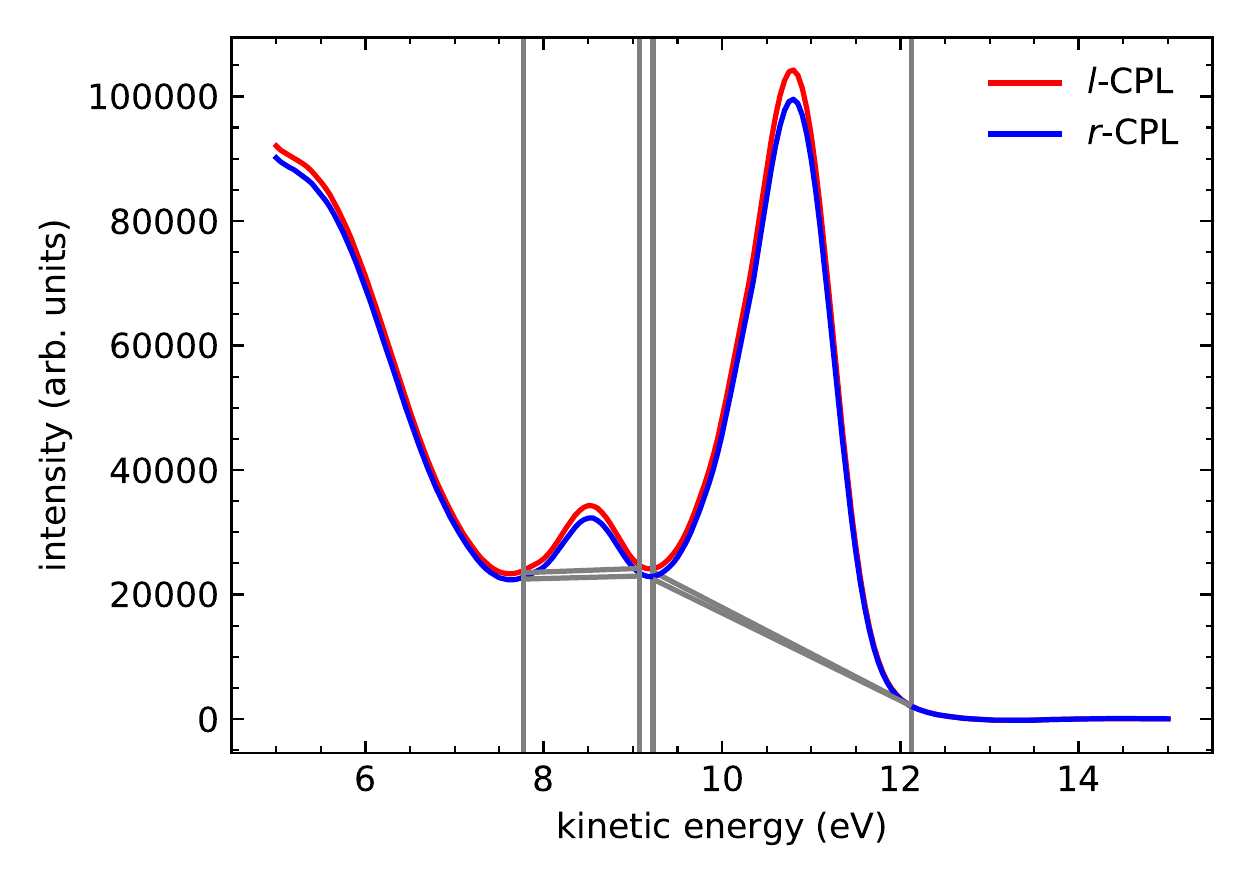}
  \caption{Photoelectron spectrum from liquid (1S,4R)-fenchone, ionized with a photon energy of 301~eV, after subtraction of an exponential background. For each photon helicity, the average over all sweeps is shown. The grey vertical lines mark regions used to determine the intensity of the C-H and C=O features, respectively, and the remaining grey lines indicate an additional background for each peak. Most of the apparent asymmetry between the two traces in this graph is an apparatus-induced artifact produced by a slight imbalance of photon intensity from the helical undulator in its two settings.}
  \label{sfig:roi}
\end{figure}
Here, we describe the peak-area integration method illustrated in Fig.~2A of the main text, termed the `roi'-method.
To determine areas of the two core-level ionization features, firstly an exponential background was subtracted individually from each sweep.
The exponential background is determined by zeroing the higher eKE end.
Averaging over the resulting data sets gives the traces shown in Supplementary Fig.~\ref{sfig:roi}.
(The traces shown in Fig.~2A of the main article differ from the results shown in Supplementary Fig.~\ref{sfig:roi} by the exclusion of a small number of outlier sweeps, and by the correction of the {\it l}-CPL trace for the apparatus asymmetry.)
In the {\it ansatz} described in connection with Fig.~2A of the main text, net areas of the peak features were then determined between a low and high kinetic energy limit as indicated, subtracting a linear background.
(Hence, in total, two different background contributions, a linear and an exponential one, were subtracted.)
Values of the chiral asymmetry parameter, $b_1$, shown in the main article were determined from these net areas, correcting for the asymmetry of the C-H line.
\subsection{Additional data on the peak-asymmetry analysis}
\begin{figure}[h!tb]
\centering
  \includegraphics[width=12cm]{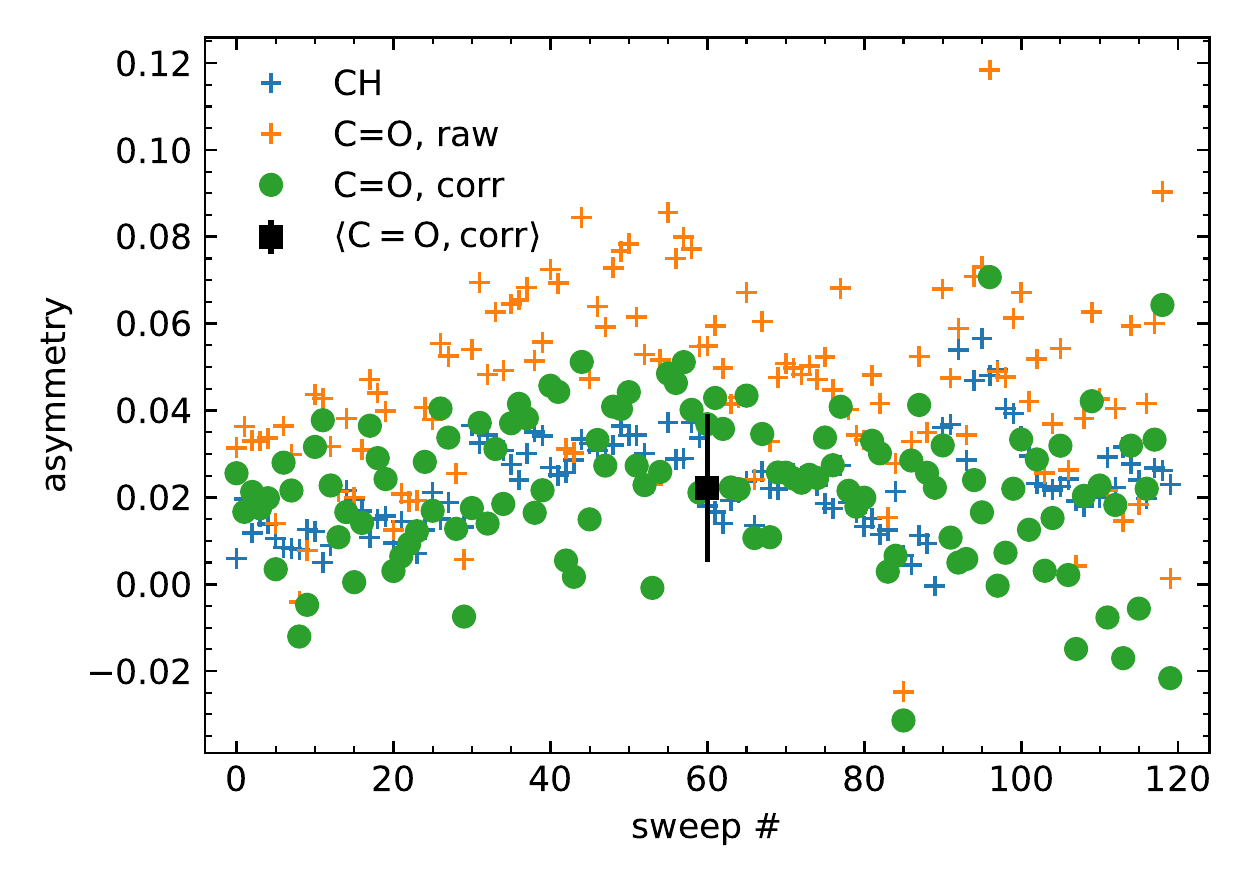}
  \caption{Asymmetry $A$ and corrected asymmetry $A_\text{corr}$ calculated from the individual sweeps of the data set shown in Supplementary Fig.~\protect\ref{sfig:traces}. The standard deviation of the set of corrected asymmetries is shown as an error bar to illustrate the data quality. The standard deviation of the mean is smaller than the symbol size. See the text for details.
  \label{sfig:asymmetry}}
\end{figure}
Here, we provide additional evidence for the robustness of our data by showing results of a slightly more expensive analysis.
From the two helicity sets of background-subtracted sweeps, we formed pairs of sweeps carried out with positive and negative photon helicity by relating them by their sweep number.
(The sweeps within a pair were not measured back-to-back, as we changed photon helicity only every thirty sweeps, in this example.)
For each pair, the asymmetry of the C-H feature and of the C=O feature were calculated, and the latter was corrected by requiring that the asymmetry of the C-H feature vanishes when the cumulative signal of the nine associated carbons is collectively considered. 
The result of this analysis is shown in Supplementary Fig.~\ref{sfig:asymmetry}.
Although the scatter between individual pairs of spectra is quite large, a clear trend prevails in the raw data: With very few exceptions, the uncorrected asymmetry of the C=O peak is larger than that of the C-H peak.
Correction of the C=O asymmetry by the apparatus asymmetry (main text, eqn.s (1)-(3)) leads to smaller values of the former, which nevertheless are significantly different from zero.
In Supplementary Fig.~\ref{sfig:asymmetry}, the standard deviation of the data set is shown for illustrative purposes. 
The standard deviation of the mean, which is a more adequate measure of the actual uncertainty of the measurement, is smaller than the symbol size (exact values for the asymmetry, standard deviation, standard deviation of the mean are 0.022, 0.017, and 0.002, respectively).
% 0.022141753773205798 0.016985685980570705 0.0015505738943763052
Supplementary Fig.~\ref{sfig:asymmetry} is based on all of the data that were recorded.
In the data analysis presented in the main article, a small number of traces that deviated in shape from the majority of the recorded sweeps were removed prior to signal averaging.
For this example, results of the asymmetry analysis were essentially unchanged from the value given above.
\subsection{Peak area integration, `exp'-method}
\label{ss:exp-method}
For the second ansatz, highlighted in the main text Fig.~2B and labeled `exp', we fitted a linear combination of an exponential and a linear function as a baseline for the averaged spectra. However, we only considered data points corresponding to the photoelectron background spectral regions in the fitting procedure. The included range is indicated in the top panel of Supplementary Fig.~\ref{sfig:exp_bkg} by the black cross markers (`fit points') which cover background regions that are well separated from the main photoemission features. The slope by which the LET ascends towards lower energies reduces at the very-low-KE end in our spectra (possibly due to the analyzer transmission function). Therefore, it is necessary to select the `fit points' such that the very-low-KE region is not included, specifically to guarantee that the main features do not intersect the zero of the x-axis. The resulting background-subtracted spectra are shown in the middle panel. The spectra were then normalized to the maximum C-H peak intensity and an 1.4~eV energy range was defined around the C=O peak, from which the asymmetries between different helicities were calculated. The result is displayed in the bottom panel of Supplementary Fig.~\ref{sfig:exp_bkg}.

\begin{figure}[h!tb]
\centering
  \includegraphics[width=15cm]{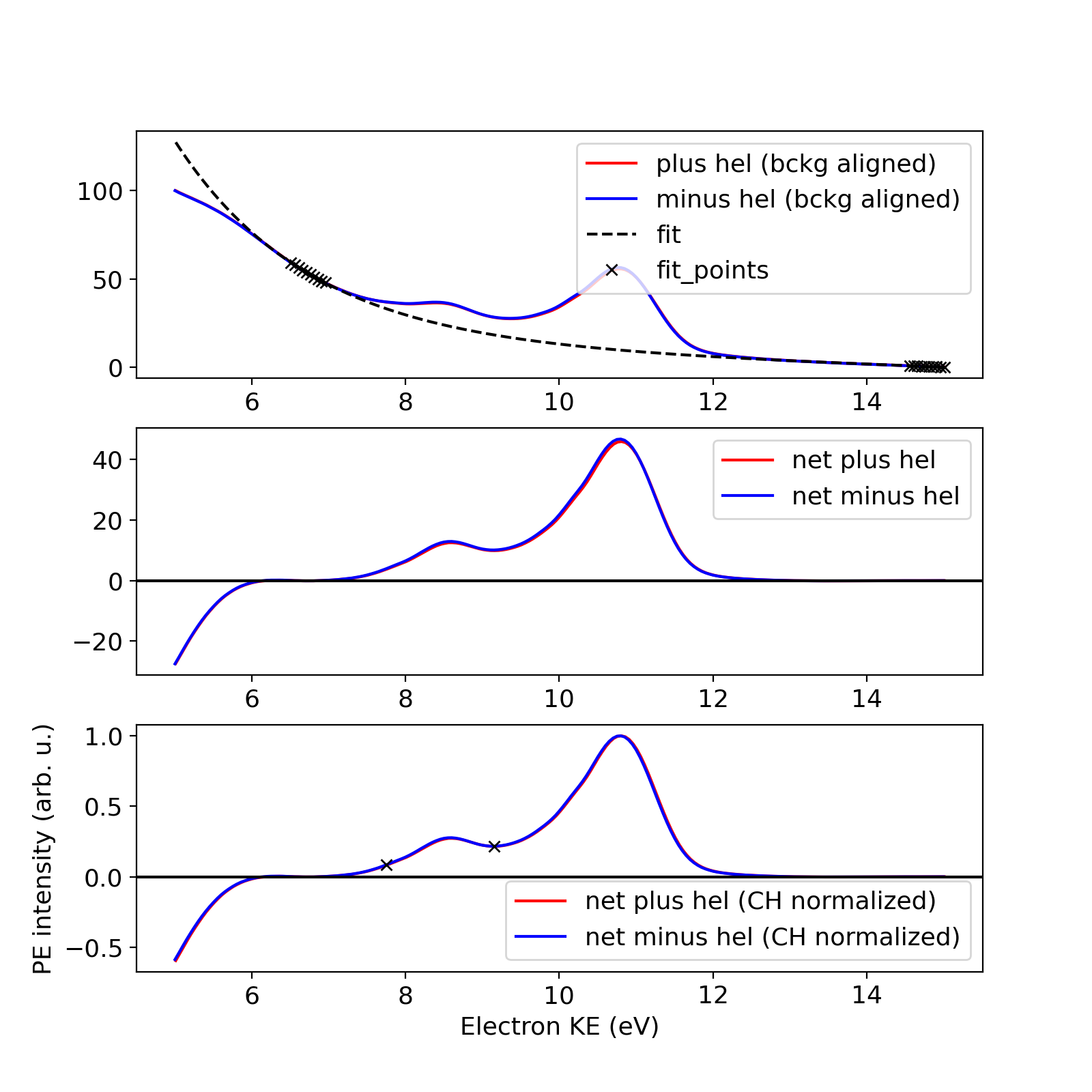}
  \caption{Background subtraction using the `exp'-method for a photoelectron spectrum from liquid (1S,4R)-fenchone, ionized with a photon energy of 301~eV. Top: For each photon helicity, the average over all sweeps is shown. The dashed line shows a background fit using a combination of a linear and an exponential function. The cross marks indicate the background signal regions used for the fit. Middle: Background-subtracted spectra. Bottom: Background-subtracted spectra, normalized to the maximum CH peak intensity at about 11~eV KE. The cross marks indicate the region used to determine the intensity of the C=O features, and from that the spectral asymmetry.}
  \label{sfig:exp_bkg}
\end{figure}
\subsection{Peak-area integration, `sum'-method}
The third ansatz, referred to in the main text Fig.~2C and labeled `sum', is similar to the method discussed above in Section~\ref{ss:exp-method}, with the only exception being that the background signal contribution was computed using a two-step approach. First, an exponential function was fitted to the high-KE end of the spectra, as displayed in the top panel of Supplementary Fig.~\ref{sfig:sum_bkg}. The resulting spectra (middle panel) where then used to perform a `total-sum fit' (also referred to as a non-iterative Shirley method\cite{Vegh2006} or total background\cite{Li}), which iterates from the high- to the low-KE end of the spectra while aggregating spectral intensities. The correspondingly created trace was then subtracted from the middle-panel spectrum, finally producing the bottom spectrum after C-H maximum-peak normalization. Note that the so-produced background spectrum was always scaled by a factor (x~0.37 in the case of Supplementary Fig.~\ref{sfig:sum_bkg}), to ensure that it did not intersect with the C 1s primary photoelectron spectrum. Thus, a final spectrum was produced that intersects with the zero of the x-axis. The background scaling factor was determined iteratively such that the background touches the spectrum at a single point only.
\begin{figure}[h!tb]
\centering
  \includegraphics[width=15cm]{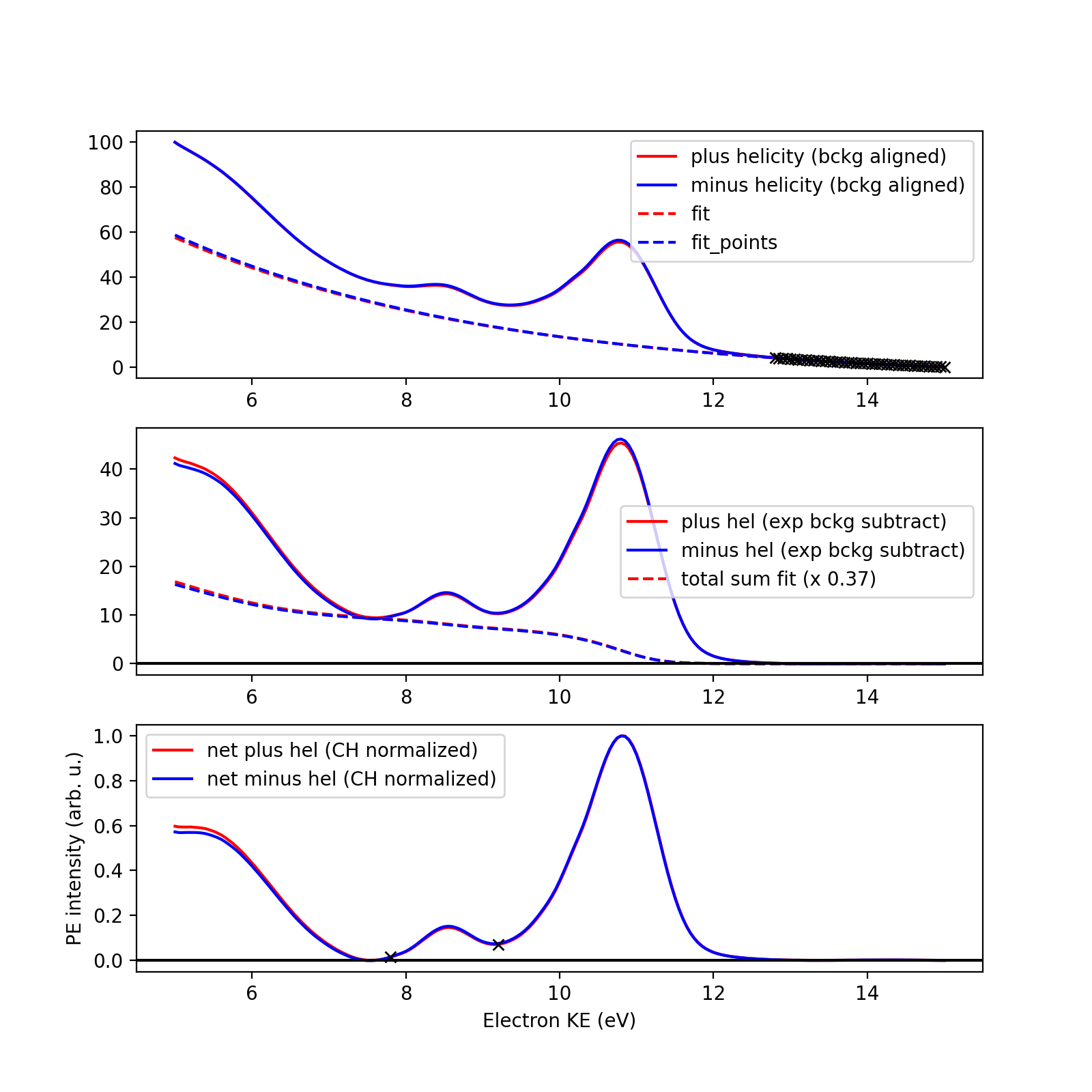}
  \caption{Background subtraction using the `sum'-method for a photoelectron spectrum from liquid (1S,4R)-fenchone, ionized with a photon energy of 301~eV. Top: For each photon helicity, the average over all sweeps is shown. The dashed line shows the background fits produced using an exponential function. The cross marks indicate regions used for the fit. Middle: Exponential background-subtracted spectra. The dashed line represents the scaled `total-sum' background. Bottom: Fully background-subtracted spectra normalized to the maximum C-H peak intensity at about 11~eV KE. The cross marks indicate the region used to determine the intensity of the C=O features.}
  \label{sfig:sum_bkg}
\end{figure}

%%%END OF MAIN TEXT%%%

%The \balance command can be used to balance the columns on the final page if desired. It should be placed anywhere within the first column of the last page.

%\balance

%If notes are included in your references you can change the title from 'References' to 'Notes and references' using the following command:
%\renewcommand\refname{Notes and references}
\renewcommand\refname{References}

%%%REFERENCES%%%
\eject   % UHe
\bibliography{supplement} %You need to replace "rsc" on this line with the name of your .bib file